\newenvironment{tablenotes}{\begin{itemize}}{\end{itemize}}

\documentclass[structabstract]{aa}  

\usepackage{graphicx}
\usepackage{txfonts}
\usepackage{natbib}
\begin{document}

   \title{Bar pattern speeds in CALIFA galaxies}

   \subtitle{I. Fast bars across the Hubble sequence}
 
   \author{J. A. L. Aguerri \inst{1,2} \and J. M\'endez-Abreu \inst{1,2} \and  J. Falc\'on-Barroso \inst{1,2} \and A. Amorin \inst{3} \and J. Barrera-Ballesteros \inst{1,2} \and 
R. Cid Fernandes \inst{3} \and R. Garc\'{\i}a-Benito \inst{4} \and B. Garc\'{\i}a-Lorenzo \inst{1,2} \and R. M.  Gonz\'alez Delgado \inst{4} \and B. Husemann \inst{5} \and V. Kalinova \inst{6} \and  M. Lyubenova  \inst{6} \and R. A. Marino \inst{7} \and I. M\'arquez \inst{4} \and D. Mast \inst{8} \and E. P\'erez \inst{4} \and S. F. S\'anchez \inst{4, 9} \and G. van de Ven \inst{6} \and C. J. Walcher \inst{5} \and N. Backsmann \inst{5} \and C. Cortijo-Ferrero \inst{4} \and J. Bland-Hawthorn \inst{10} \and  A. del Olmo \inst{4} \and J. Iglesias-P\'aramo \inst{4,9}  \and I. P\'erez \inst{11,12} \and P. S\'anchez-Bl\'azquez \inst{13} \and L. Wisotzki \inst{5} \and B. Ziegler \inst{14}}

   \institute{Instituto de Astrof\'{\i}sica de Canarias. C/ V\'{\i}a L\'actea s/n, 38200 La Laguna, Spain \label{inst1} \and
Departamento de Astrofísica, Universidad de La Laguna, 38205 La Laguna, Tenerife, Spain \label{inst2} \and Departamento de Fisica, Universidade Federal de Santa Catarina. P.O. Box 476, 88040-900. Florian\'opolis, S C, Brazil \label{inst3} \and
 Instituto de Astrofísica de Andalucía - CSIC, 18008, Granada, Spain \label{inst4} \and Leibniz-Institut für Astrophysik Potsdam (AIP), An der Sternwarte 16, D-14482 Potsdam, Germany \label{inst5}  \and Max-Planck-Institut für Astronomie, Konigstuhl 17, 69117 Heidelberg, Germany \label{inst6} \and CEI Campus Moncloa, UCM-UPM, Departamento de Astrofísica y CC. de la Atmósfera, Facultad de CC. Físicas, Universidad Complutense de Madrid, Avda. Complutense s/n, 28040 Madrid, Spain \label{inst7} \and Instituto de Cosmologia, Relatividade e Astrof\'{i}sica – ICRA, Centro Brasileiro de Pesquisas F\'{i}sicas, Rua Dr. Xavier Sigaud 150, CEP 22290-180, Rio de Janeiro, RJ, Brazil \label{inst8} \and Centro Astron\'omico Hispano Aleman, Calar Alto (CSIC-MPG), C/ Jes\'us Durb\'an Rem\'on 2-2. E-04004 Almer\'{\i}a, Spain \label{inst9}    \and Sydney Institute for Astronomy. School of Physics A28. University of Sydney. NSW 2006, Australia  \label{inst10} \and Dpto. de Física Teórica y del Cosmos, University of Granada, Facultad de Ciencias (Edificio Mecenas), 18071, Granada, Spain  \label{inst11} \and Instituto Universitario Carlos I de Física Teórica y Computacional, Facultad de Ciencias, 18071, Granada, Spain \label{inst12} \and Departamento de F\'{\i}sica Te\'orica, Universidad Autónoma de Madrid, Cantoblanco, E-28049 Madrid, Spain \label{inst13} \and University of Vienna, Türkenschanzstrasse 17, 1180, Vienna, Austria \label{inst14}
             }

   \date{Received ; accepted }

 
  \abstract
  {The bar pattern speed  ($\Omega_{\rm b}$) is defined as the rotational frequency of the bar, and it determines the bar dynamics. Several methods have been proposed for measuring $\Omega_{\rm b}$. The non-parametric method proposed by Tremaine \& Weinberg (1984; TW) and based on stellar kinematics is the most accurate. This method has been applied so far to 17 galaxies, most of them SB0 and SBa types. }
   {We have applied the TW method to a new sample of 15 strong and bright barred galaxies, spanning a wide range of morphological types from SB0 to SBbc.  Combining our analysis with previous studies, we investigate 32 barred galaxies with their pattern speed measured by the TW method. The resulting total sample of barred galaxies allows us to study the dependence of $\Omega_{\rm b}$ on galaxy properties, such as the Hubble type.}
   {We measured $\Omega_{\rm b}$  using the TW method on the stellar velocity maps provided by the integral-field spectroscopy data from the CALIFA survey. Integral-field data solve the problems that long-slit data present when applying the TW method, resulting in the determination of more accurate $\Omega_{\rm b}$. In addition, we have also derived the ratio $\cal{R}$ of the corotation radius to the bar length of the galaxies. According to this parameter, bars can be classified as fast ($\cal{R}$ $< 1.4$) and slow ($\cal{R}$>1.4). }
   {For all the galaxies, $\cal{R}$ is compatible within the errors with fast bars. We cannot rule out (at 95$\%$ level) the fast bar solution for any galaxy.  We have not observed any significant trend between $\cal{R}$ and the galaxy morphological type.  }
   {Our results indicate that independent of the Hubble type, bars have been formed and then evolve as fast rotators. This observational result will constrain the scenarios of formation and evolution of bars proposed by numerical simulations. }


\titlerunning{Pattern speed in CALIFA barred galaxies}
\authorrunning{J. A. L. Aguerri et al.}

   \maketitle
%

\section{Introduction}

Bars are ellipsoidal-like structures present in the inner  region of most disc galaxies. The galaxy classification into barred and non-barred has been made since the pioneering Hubble morphological classification. Indeed, barred galaxies constitute one of the branches of the so-called Hubble tuning-fork diagram \citep[see ][]{hubble1936}. 

About 40-50$\%$ of nearby disc galaxies observed in the optical wavelengths have been classified as  barred systems \citep[see, e.g., ][]{marinova2007,barazza2008,aguerri2009}. This fraction is even higher (about 60-70$\%$) when galaxies are observed through near-infrared filters \citep[see, e.g., ][]{knapen2000,eskridge2000,menendezdelmestre2007}.  The bar fraction depends on different galaxy properties, such as Hubble type, mass, and environment. There is a debate in the literature about the dependence of the bar fraction on the Hubble type \citep[e.g.,][]{aguerri2009,buta2010, barway2011,masters2011,marinova2012}.  Several studies have recently demonstrated that total stellar mass is an important physical parameter that regulates bar formation \citep{mendezabreu2010,nair2010,sheth2008}. However, the environment also plays a non-negligible role in their formation and evolution, as shown by Mendez-Abreu et al. (2012) when comparing the bar fraction in three different environments (field, Virgo, and Coma clusters). They show that the fraction of bars in the Virgo and Coma clusters are statistically different from the field one. They argue that  bright disc galaxies are stable enough against interactions, whereas for fainter galaxies, interactions in cluster environment become strong enough to heat up the discs, inhibiting bar formation and even destroying the discs \citep[see also][]{sanchezjanssen2010}. 

Bars are very prominent features in light. In contrast, they represent a small fraction of the mass of the discs of galaxies. Hydrodynamical simulations of barred galaxies show that the bar component contributes only about 10\% to 20$\%$ of the disc potential \citep[see][]{england1990b, laine1999, lindblad1996a, weiner2001, aguerri2001}. Nevertheless, they produce important changes in the dynamics of the discs when they are present. In fact, simulations also show that bars are very efficient at redistributing angular momentum, energy, and mass on both luminous and dark matter galactic components \citep[see][]{weinberg1985,debattista1998,debattista2000,athanassoula2003,martinezvalpuesta2006,sellwood2006a,sellwood2006b,villavargas2009}. 

Bars are fully characterized by three parameters: length, strength, and pattern speed. Both length and strength  can be determined  using optical and/or near-infrared images. In contrast, the bar pattern speed is a dynamical parameter, and its determination requires kinematics.

The length of the bars has been determined by optical visual inspection \cite{kormendy1979,martin1995}, locating the maximum of the isophotal ellipticity \cite{wozniak1995, marquez1999,laine2002, marinova2007, aguerri2009}, or by structural decompositions of the galaxy surface brightness distribution \cite{prieto1997,prieto2001,aguerri2001,aguerri2003,aguerri2005,laurikainen2005,laurikainen2007,gadotti2008, laurikainen2009, weinzirl2009,gadotti2011}. All these studies have shown that a typical bar radius is 3-4 kpc \cite[e.g.,][]{marinova2007,aguerri2009}. There is a clear dependence between the bar length and other galaxy parameters, such as  disc scale length, galaxy size, galaxy colour, or prominence of the bulge \cite{aguerri2005,marinova2007,gadotti2011,hoyle2011}. In addition, there is a dependency of the bar length on the Hubble type. In particular, S0 galaxies show larger bars than late-type ones \citep[][but see also Masters et al. 2011]{elmegreen1985,aguerri2009,erwin2005,menendezdelmestre2007}.

The bar strength is a parameter that measures the non-axisymmetric forces produced by the bar potential in the disc of galaxies \citep[see, e.g.,][]{laurikainen2002}. The most popular methods for determining the bar strength are to measure the torques of the bar \citep[e.g.,][]{combes1981,quillen1994,buta2001,laurikainen2007,salo2010}, bar ellipticity \cite{martinet1997,aguerri1999,whyte2002,marinova2007,aguerri2009}, or Fourier decomposition of the galaxy light \cite{ohta1990,marquez1996,aguerri2000,athanassoula2002,laurikainen2005}. In general, S0 galaxies show weaker bars than late-type ones \cite[see, e.g.,][]{laurikainen2007,aguerri2009,buta2010}.

The bar pattern speed ($\Omega_{\rm b}$)  is a pure dynamical parameter that fully determines the dynamics of a barred galaxy. It is defined as the rotational frequency of the bar. Theories of stellar orbits have shown that $\Omega_{\rm b}$ has a physical upper limit. Thus,  a bar cannot extend beyond the corotation  resonance (CR) radius ($R_{\rm CR}$) of the galaxy. This corotation radius is the region of the galaxy where the angular speed of the stars of the disc in circular motions equals the bar pattern speed. This limit is imposed by the stability of the main family of orbits forming the bar (the so-called $x_{1}$ family; Contopoulos 1980) which are only stable within $R_{\rm CR}$. Orbits in the outer region of the disc ($R>R_{\rm CR}$) cannot support a bar structure \citep[][]{contopoulos1980,athanassoula1992}.

Measuring the bar pattern speed is technically the most difficult amongst the observational parameters characterising bars, and several methods have been developed in the literature to determine it. Hydrodynamical simulations of individual galaxies have been extensively used to determine $\Omega_{\rm b}$ in barred galaxies.  These methods start by computing the gravitational potential of the galaxy that is used to produce hydrodynamical galaxy models, with $\Omega_{\rm b}$ as one of the free parameters of the models. Determination of the best value of $\Omega_{\rm b}$ is done by matching the modelled and observed surface gas distribution and/or gas velocity field  \citep[see, e.g.,][]{sanders1980,hunter1988,england1990,garciaburillo1993,sempere1995a,lindblad1996a,
lindblad1996b,laine1999,weiner2001,aguerri2001,perez2004,rautiainen2008,treuthardt2008}. Other methods determine the pattern speed by identifying some morphological galaxy features with Lindblad resonances. We can mention the position of galaxy rings \citep[see, e.g.,][] {buta1986,buta1995,vegabeltran1997,munoztunon2004,perez2012}; changes in the morphology or phase of spiral arms near $R_{\rm CR}$ \citep{canzian1993,canzian1997,puerari1997,aguerri1998,buta2009}; detecting the offset and shape of dust lanes \citep[see, e.g.,][]{vanalbada1982,athanassoula1992}; locating colour and star formation changes outside the bar region \cite[][]{cepa1990,aguerri2000}; or studying the morphology of the residual gas velocity field after the rotation velocity subtraction \citep{sempere1995b,font2011,font2014}. Nevertheless, the most accurate method for measuring the bar pattern speed is the non-parametric method proposed by Tremaine \& Weinberg (1984; hereafter TW). The goal of this paper is to apply the TW method to a large sample of barred galaxies.

The organization of the paper is as follows. The theoretical basis of the TW method is shown in Sect. 2. Section 3 shows the description of the galaxy sample. Section 4 presents the photometric parameters of the bars. The stellar velocity maps of the galaxies are shown in Sect. 5, and results are given in Sect. 6. The discussion and conclusions are shown in Sects. 7 and 8, respectively. Throughout this paper we have used the cosmology $H_{0}=70$ km Mpc$^{-1}$ s$^{-1}$, $\Omega_{m}=0.3$, and $\Omega_{\lambda}=0.7$.

\section{Theoretical basis of the TW method}

We let ($X, Y$) be a Cartesian coordinate system in the sky plane and take the origin at  the centre of the galaxy, with $X$-axis coincident with the line of nodes (LON), which is defined as the intersection between the sky plane and the plane of the disc of the galaxy. Assuming that the disc of the galaxy has a well-defined pattern speed $\Omega_{\rm b}$ and that the surface brightness of the measured tracers follow the continuity equation, the TW method is based on the equation:

\begin{equation}
\Omega_{\rm b} \sin i = \frac{\int_{-\infty}^{+\infty} h(Y) \int_{-\infty}^{+\infty} \Sigma(X,Y)\,V_{LOS}(X,Y)\,dX\,dY}{\int_{-\infty}^{+\infty} h(Y) \int_{-\infty}^{+\infty} X\,\Sigma(X,Y)\,dX\, dY} \equiv \frac{\langle V \rangle}{\langle X \rangle},
\end{equation}

where $i$ is the galaxy inclination, $V_{LOS}(X,Y)$ is the line of sight (LOS) velocity, $\Sigma(X,Y)$ represents the surface brightness of the galaxy, and $h(Y)$ is a weight function. The numerator of the previous equation represents the mean velocity weighted by the light of the galaxy ($\langle V \rangle$). The denominator shows the weighted mean position of the tracers ($\langle X \rangle$). We denote the numerator and denominator of Eq. (1) as "kinematic" and "photometric" integrals, respectively.  All previous parameters can be measured from photometric or kinematic observations. Thus, $i$ can be obtained from the ellipticity of the outermost galaxy isophotes, $\Sigma(X,Y)$ can be determined from optical photometry, and the velocity field of the galaxy $(V_{\rm LOS}(X,Y))$ can be measured from long-slit or integral-field spectroscopy. In practice, the integrals from Eq. (1) are calculated along several directions oriented parallel to the LON but offset by a distance $Y_{0}$. The slope of the straight line defined by the $\langle V \rangle$ versus $\langle X \rangle$ computed points represents $\Omega_{\rm b} \sin i$ \citep[see][]{tremaine1984}. In the case of long-slit spectroscopy, the weight function is given by $h(Y)=\delta (Y-Y_{0})$. Similar weight functions can be defined for integral-field spectroscopy.  Formally, integrals in Eq. 1 are over $-\infty < X, Y < +\infty$. Nevertheless, they can be limited to a finite value of $X$ and $Y$ if the axisymmetric part of the disc is reached (see Sect. 6.1). 

Stellar kinematics has been used to measure $\Omega_{\rm b}$ using the TW method for a  large sample of early-type galaxies \citep[e.g.,][]{kent1987,merrifield1995,gerssen1999,debattista2002,aguerri2003,corsini2003,debattista2004,corsini2007}. In contrast, few pattern speeds have been determined through this method for late-type galaxies \citep[e.g.,][]{gerssen2003,treuthardt2007}. See also Corsini (2011) for a review of the bar pattern speed measurements using the TW method. The lack of measurements of bar pattern speed in late-type galaxies by the TW method is basically due to the effects produced on the determination of $\Omega_{\rm b}$ by the uncertainties in the  position angle (hereafter PA), $i$,  and/or the effect of the presence of  dust and star formation in late-type discs. Debattista (2003) demonstrates that  errors of a few degrees in determining the PA of galaxies can significantly change the measured value of $\Omega_{\rm b}$ using the TW method. For this reason galaxies with small errors in PA and/or $i$ should be selected. In addition, the measured value of $\Omega_{\rm b}$ can also be affected by dust and/or star formation in late-type galaxies, since in these cases the light does not trace the mass distribution of these galaxies.  Nevertheless, these effects can be mitigated by computing the kinematic and photometric integrals of Eq. (1) using the mass distribution as weight \citep[see][]{gerssen2007}.

In recent years, an extension of the TW method explained above has been applied to galaxies that show multiple pattern speeds \citep[see][]{maciejewski2006,corsini2007,meidt2009}. The TW method was also  applied to gas tracers, such as CO \citep[][]{zimmer2004,rand2004} or H$\alpha$ \citep[][]{hernandez2005, emsellem2006, fathi2007, chemin2009, gabbasov2009, fathi2009}.

\subsection{Fast and slow bars}

Barred galaxies are commonly classified according to the distance-independent ratio ${\cal R}=R_{\rm CR}/a_{\rm b}$, where $R_{\rm CR}$ and $a_{\rm b}$ are the corotation and bar radius, respectively. Theoretical works based on stellar orbits in barred potential predict that ${\cal R}=1.2\pm0.2$ \citep[see][]{athanassoula1992}. This parametrization permits a classification of bars into fast ($1.0<{\cal R}<1.4$) and slow ones (${\cal R}>1.4$). Most  observed bars have turned out to be fast bars; nevertheless, there are few bars in the literature compatible with being slow bars \citep[see][]{rautiainen2008}.

There are some hints in previous works that late-type galaxies could have higher mean values of ${\cal R}$ than early-type ones \citep[see][]{aguerri1998}. More recent results based on hydrodynamical models and comparing the morphology of real galaxies with models show that while early-type galaxies always have fast bars, late-type galaxies host both slow and fast bars \citep[see][]{rautiainen2008}. Nevertheless, this result has not been confirmed in a significant sample of galaxies of different morphological types by using the TW method. The sample of late-type galaxies analyzed so far by the TW is very small in order to infer some dependence between ${\cal R}$ and the Hubble type. The aims here are to determine the pattern speed of a large sample of galaxies throughout the Hubble sequence by using the TW method and to study the dependence of the pattern speed of the bar with the Hubble type.


\section{The CALIFA sample of barred galaxies}

The barred galaxies used in the present study were taken from the Calar Alto Legacy Integral Field Area (CALIFA) Survey\ \citep[][]{sanchez2012}. CALIFA's mother sample is formed by 939 galaxies selected from the SDSS-DR7 photometric catalogue \citep[][]{abazajian2009}. The main selection criteria were angular isophotal diameter $45 < D_{25} <80$ arcsec and  redshift range $0.005 < z < 0.03$. These criteria ensure that the selected objects fit well into the field of view (FOV) of the instrument. This survey aims to obtain spatially resolved spectroscopic information for a fraction of the mother-galaxy sample ($\approx 600$ galaxies), limited by available telescope time. For more details about the properties of the galaxies of the CALIFA mother sample, see Walcher et al. (2014). This project has been the biggest effort with integral field spectroscopy so far.

The CALIFA observations were carried out at the 3.5m telescope of the Calar Alto observatory with the Potsdam Multi Aperture Spectrograph \citep[PMAS;][]{roth2005} in PPAK mode. This instrumental mode consists of 382 fibres of 2.7 arcsec diameter each. These fibres cover a FOV of 74$'' \times 64 ''$. A dithering scheme of three pointings were adopted to cover the full FOV. This allows us to have a final resolution of 1 arcsec \citep[see][]{sanchez2007,perezgallego2010,rosalesortega2010,sanchez2012}. Both the FOV and the spatial resolution of the observations make this dataset ideal for studies of extended disc galaxies proposed here. 

The objects were observed using two different setups. First, the grating called V500 shows a nominal resolution of $\lambda/\Delta \lambda = 850$ at 5000 $\AA$ and covers from 3745 to 7300 $\AA$. The second setup was done using the V1200 grating with better spectral resolution $\lambda/\Delta \lambda = 1650$ at 4500 $\AA$. This grating covers from 3400 to 4750 $\AA$. In the present work we used the observations throughout the V1200 grating. It was selected in order to have reliable velocity dispersion curves of the galaxies needed for computing $R_{CR}$ (see Sect. 5.2). The galaxy sample of the present paper was selected among the 200 that were first observed and reduced objects with the V1200 grating. For more details about the observation strategy and the data reduction process, see S\'anchez et al. (2012) and Husemann et al. (2013). 

The CALIFA mother-sample photometric properties were obtained from SDSS-DR7. In addition, the galaxies of the CALIFA mother sample were morphologically classified by using the $i$-band SDSS images by a group of five people from the CALIFA team \citep[see][]{walcher2014}. The total number of galaxies showing strong bars (SB) on the CALIFA sample turned out to be 156. Only 41 of them have already been observed with the V1200 grating. Nevertheless, the TW method cannot be applied to all galaxies. Galaxy inclination, position angle, and bar orientation should be considered for its application. In particular, low and high inclined galaxies and those with bars near the galaxy's major or minor axis should be not taken into account because the TW is not applicable to these objects. Therefore, we excluded from our analysis all bars that show PA less than 10$^{\circ}$ with respect to the major or minor axis of the galaxies and either those face-on ($b/a>0.86$) or edge-on ($b/a<0.34$) galaxies. This restriction reduces the total number of barred galaxies to 20. In addition, five more galaxies were excluded for other reasons, such as the  large number of field stars, the non-flat rotation curve at large radius, or/and the insufficient quality of the stellar velocity maps. 

The final selected sample of barred galaxies contains objects that cover the morphological types: 1 SB0, 2 SB0/a, 1 SBa, 1 SBab, 6 SBb, and 4 SBbc. Similar to the CALIFA mother sample, our galaxy sample is dominated by  spirals (SBb-SBbc). This is appropriate for filling the gap of pattern speeds measured by the TW method in late-type galaxies. 

\begin{figure}
   \centering
   \includegraphics[angle=0,scale=0.5]{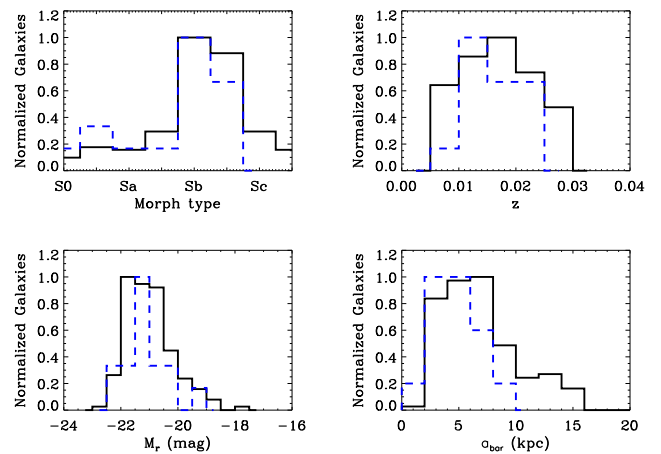}
   \caption{ Normalized distribution of morphological types (a), redshift (b), absolute $r$-band magnitudes (c), and bar radius (d) of the CALIFA's mother-sample barred galaxies (full black line histograms) and the barred galaxies selected in the present paper (dashed blue-line histograms).}
              \label{comp_sample}%
    \end{figure}

Figure \ref{comp_sample} shows the comparison of the distribution of morphological types, redshift, absolute $r$-band magnitudes, and bar radius between the barred galaxies in the CALIFA mother sample and the selected barred galaxies of the present work.  Our sample is a small subsample of the total bar population of the CALIFA mother sample. Nevertheless, the Kolmogorov-Smirnov test indicates that our subsample does not have a statistically different distribution of morphological types, redshift, and absolute magnitudes from the total sample of barred galaxies.  The actual sample loss bars have radii larger than 10 kpc. The lack of large bars is due to: i) galaxy inclination, ii) bar orientation, and/or iii) no observations using the V1200 grating. We never cut the bar sample according to bar size. It is also worth noticing that the galaxies presented here are all bright galaxies ($M_{r}>-19.0$). In this range of magnitudes, CALIFA is representative of the local Universe. No dwarf galaxies have been considered in the present work. The $r-$band SDSS images of the final sample are shown in Fig. \ref{f1} and the main parameters of the galaxies are given in Table \ref{tb1}.

In summary, the sample presented in this work overcomes two major problems of other samples presented in the literature. First, integral-field observations such as those presented here get rid of many of the problems of long-slit observations. Second, our sample is dominated by late-type galaxies, the missing piece of information in the TW studies of bar pattern speeds.

\section{Photometric parameters of the barred galaxies}

\subsection{Inclinations and position angles}

Analysis of the galaxy isophotes provides the inclinations\footnote{The inclination ($i$) of a disc is given by $b/a=(q^{2}+(1-q)^{2} \times cos^{2}(i))^{1/2}$, where $b/a$ is the ratio between the observed minor and major axis of the disc, and $q$ represents its intrinsic thickness. In this paper we have assumed the thin disc approximation ($q=0$). Thus, we have computed the inclination of the galaxies by $cos(i)=b/a$}, PA, and other important information about the different structural components of galaxies \citep[see, e.g.,][]{wozniak1995,aguerri2000b}. The galaxy inclination makes the isophotes of the galaxies appear as ellipses in the plane of the sky.  We fitted the isophotes of our galaxies by ellipses using the ELLIPSE routine from the IRAF package \citep[][]{jedrzejewski1987}. The ellipticity and PA isophotal radial profiles obtained from these fits are shown in Fig. \ref{f1_phot}.

The ellipticity radial profile of an unbarred spiral galaxy grows from almost zero values at the centre of the object up to a constant value at large radii, which corresponds to the galaxy inclination. The PA isophotal radial profile also reaches a constant value at large radii, which corresponds to the orientation of the LON of the galaxy. This behaviour can be observed in the ellipticity and PA isophotal radial profiles shown in Fig. \ref{f1_phot}. We therefore computed the $i$ and PA of our barred galaxies by averaging their outer isophotes. The results are given in Table \ref{tb1}.

  \begin{table*}
      \caption[]{Main parameters of the CALIFA barred galaxy sample}
         \label{tb1}
         \centering
         \begin{tabular}{c c c c c c c c}
            \hline
            \hline
           Galaxy & RA (J2000)  & DEC (J2000) & Morph. Type & $i$ & PA & $R_{e}$ & $m_{r}$ \\
                     &   (hh:mm:ss) & ($^{o}: ': ''$) &    & (degrees) & (degrees) & (arcsec) & (mag)\\
            \hline 
            NGC~0036 & 00:11:22.3 & 06:23:22  & SBb  & 57.2$\pm$3.7 & 23.4 $\pm$1.3 & 14.9 & 12.7 \\ 
            NGC~1645 & 04:44:06.4 & -05:27:56 & SB0a & 64.5$\pm$0.6 & 84.7 $\pm$0.4 & 9.2 & 12.7\\
            NGC~3300 & 10:36:38.4 & 14:10:16  & SB0a & 57.2$\pm$0.7 & 172.0$\pm$0.3 & 12.0 & 12.3\\
            NGC~5205 & 13:30:03.6 & 62:30:42  & SBbc & 50.0$\pm$0.7 & 170.1$\pm$1.6 & 15.9 & 12.5\\
            NGC~5378 & 13:56:51.0 & 37:47:50  & SBb  & 37.8$\pm$3.1 & 86.5 $\pm$5.4 & 19.9 & 12.5\\
            NGC~5406 & 14:00:20.1 & 38:54:56  & SBb  & 44.9$\pm$0.4 & 111.8$\pm$0.8 & 18.1 & 12.2\\
            NGC~5947 & 15:30:36.6 & 42:43:02  & SBbc & 44.6$\pm$1.4 & 72.5 $\pm$3.1 & 12.4 & 13.4\\
            NGC~6497 & 17:51:18.0 & 59:28:15  & SBab & 60.9$\pm$0.7 & 112.0$\pm$0.4 & 10.7 & 13.0\\
            NGC~6941 & 20:36:23.5 & -04:37:07 & SBb  & 42.3$\pm$0.8 & 127.5$\pm$2.7 & 14.3 & 13.1\\ 
            NGC~6945 & 20:39:00.6 & -04:58:21 & SB0  & 51.3$\pm$2.3 & 126.1$\pm$1.3 & 9.3 &12.4\\
            NGC~7321 & 22:36:28.0 & 21:37:19  & SBbc & 48.3$\pm$0.5 & 13.4 $\pm$1.3 & 12.0 & 12.9\\
            NGC~7563 & 23:15:55.9 & 13:11:46  & SBa  & 55.8$\pm$2.4 & 149.8$\pm$1.3 & 8.1 & 12.4\\
            NGC~7591 & 23:18:16.3 & 06:35:09  & SBbc & 57.6$\pm$0.5 & 144.0$\pm$2.0 & 12.0 & 12.6\\
           UGC~03253 & 05:19:41.9 & 84:03:09  & SBb  & 56.8$\pm$1.6 & 92.0 $\pm$1.7 & 11.8 & 13.2\\
           UGC~12185 & 22:47:25.0 & 31:22:25  & SBb  & 64.0$\pm$0.6 & 161.0$\pm$1.0 & 8.8 & 13.5\\
            \hline
        \end{tabular}
\begin{tablenotes}
\small
\item Note: Columns are: (1) NGC name of the galaxy; (2) galaxy right ascension, (3) galaxy declination, (4) morphological type, (5) galaxy inclination measured from isophotal ellipticity profiles on i-band SDSS images, (6) galaxy PA measured from isophotal PA radial profiles on i-band SDSS images, (7) effective radius in r-band from SDSS-DR9, (8)  model r-band magnitude from SDSS-DR9
\end{tablenotes}
   \end{table*}

\subsection{Determination of the bar radius}

The determination of the bar radius is not an easy task. Several methods have been proposed in the literature during the past few decades. We used three of the most popular methods ($a_{\rm b,1}, a_{\rm b,2},$ and $a_{\rm b,3}$) for estimating the bar radius ($a_{\rm b}$) of our galaxies. 

Two of the bar radius measurements ($a_{\rm b,1}$ and $a_{\rm b,2}$) were obtained by using the information of the ellipticity and PA isophotal radial profiles and the different methods proposed in the literature \citep[see][]{marquez1999,athanassoula2002,micheldansac2006,aguerri2009}. These methods are based on the peculiar features produced by the shape and orientation of the stellar orbits of barred galaxies \citep[][]{contopoulos1980,athanassoula1992}. In particular, the galaxy isophotes are almost circular at the galaxy centre, because of either seeing  effects or the presence of a  spherical bulge. As we get  away  from  the  centre,  there  is a  general  increase  in the ellipticity up to a local maximum, and then it suddenly decreases towards a  minimum at  the location  where the  isophotes  become axisymmetric (disc region)  in the face-on  case.  Typical measurements of  the bar radius using  the ellipticity profile involve  measuring the position of  this  maximum  and  minimum. In fact, they represent two extreme  cases  \citep[][]{micheldansac2006},  and therefore they can be understood as the lower and upper limits to the bar radius measurement. We adopted the position of the local maximum of the isophotal ellipticity of our galaxies as one measurement of the bar radius of
our galaxies. Table \ref{tb2} shows the values of the bar radius for our galaxies $a_{\rm b,1}$ determined by this method. 

The second estimate of the bar radius ($a_{\rm b,2}$) takes the information provided by the isophotal radial PA profiles into account. The PA radial profile  is also characteristic of barred galaxies, since it is constant in the bar region and then changes to fit the outer disc orientation \citep[e.g.,][]{wozniak1995,aguerri2000b}. A  typical bar length  is measured  at the  radius where  the position angle changes by $\Delta$PA with respect to  the value corresponding to the maximum ellipticity. Usually $\Delta$PA=5$^{\circ}$  is a good value for the bar radius \citep[see e.g.,][]{aguerri2009}. The value of the bar radius obtained with this method could be correlated to $a_{b,1}$. This is due to the behaviours inside the bar region of the ellipticity, and PA profiles are similar among galaxies. The values of $a_{\rm b,2}$ for our galaxies can be seen in Table \ref{tb2}.

Fourier decomposition has been used extensively in characterising structures, like bars, which represents a bisymmetric departure from axisymmetry. We used the method based on Fourier decomposition of the light distribution of the galaxies proposed by Aguerri et al. (2000). Following this method, the bar radius  is determined by the ratios of the intensities in the bar and inter-bar regions. The azimuthal surface brightness profiles of the deprojected galaxies were decomposed in a Fourier series. The bar intensity, $I_{\rm b}$, is defined as $I_{\rm b}=I_{0}+I_{2}+I_{4}+I_{6}$ (where $I_{0}, I_{2}, I_{4}$ and $I_{6}$ are the $m=0, 2, 4,$ and 6 terms of the Fourier decomposition, respectively). Similarly, the inter-bar intensity is defined as $I_{\rm ib}=I_{0}-I_{2}+I_{4}-I_{6}$. The bar region is defined as the region where $I_{\rm b}/I_{\rm ib} > 0.5\times[max(I_{\rm b}/I_{\rm ib})-min(I_{\rm b}/I_{\rm ib})]+min(I_{\rm b}/I_{\rm ib})$. The bar radius ($a_{b,3}$) is identified as the outer radius at which $I_{\rm b}/I_{\rm ib} > 0.5\times[max(I_{\rm b}/I_{\rm ib})-min(I_{\rm b}/I_{\rm ib})]+min(I_{\rm b}/I_{\rm ib})$. Numerical simulations have shown that this method determines the bar radius within 8$\%$ accuracy except for very thin bars \cite[see, e.g.,][]{athanassoula2002}. The values of $a_{\rm b,3}$ for our sample are shown in Table \ref{tb2}. 

Aguerri et al. (2009) demonstrated that the best method for determining the bar radius depends on the shape of the surface brightness profile of the bar (see their fig. 4). This bar type is only available after accurate multi-component surface-brightness decomposition of photometrical images. They also demonstrated that $a_{\rm b,1} < a_{\rm b,2} < a_{\rm b,3}$ independent of the bar type. In addition, the real bar radius is always between $a_{\rm b,1}$ and $a_{\rm b,3}$. For these reasons, we consider $a_{\rm b}$ as the mean of $a_{\rm b,1},  a_{\rm b,2}$, and $ a_{\rm b,3}$. Moreover, the upper and lower uncertainties of the bar radius are the differences between $a_{\rm b}$ and $a_{\rm b,1},  a_{\rm b,2}$, and $ a_{\rm b,3}$, respectively. 

\section{Stellar velocity maps}

The stellar kinematics of the galaxies was measured from the spectral datacubes observed with the V1200 grating. The full description of the procedure will be explained in Falcon-Barroso et al. (in prep.). For the sake of clarity, we briefly summarise the process here. 

In the first step, the spaxels of the datacube were  Voronoi-binned to achieve a limiting signal-to-noise ratio $S/N > 20$ \citep[see][]{cappellari2003}, while spectra with $S/N<3$ were not considered. The values of the LOS velocity and velocity dispersion were obtained by fitting the binned spectra using the penalised pixel-fitting method (pPXF) from Cappellari \& Emsellem (2004). The fit takes the continuum and the galaxy absorption features presented in the wavelength range of the spectra into account. Emission lines were masked where present in the spectra of the galaxies. A non-negative linear combination of a subset of 328 stellar templates from the Indo-US library \citep[][]{valdes2004} were used for the fit of the spectra. This subset was carefully selected in the parameter space defined by the stellar properties: $T_{eff}$, log($g$), and  [Fe/H]. Errors in both velocities and velocity dispersion were obtained via Monte Carlo simulations.

The stellar velocity maps of the galaxies considered in the present paper are shown in Fig \ref{f2}. In these maps, the systemic velocity of the galaxies was computed as the average velocity of the stars in the central 5$''$ aperture and subtracted from the velocity maps. The stellar velocity obtained for our galaxies was used for computing the kinematic integrals of the TW method (see Eq. 1). Figure \ref{f3} shows stellar-streaming mean velocities with $V_{sys}$ subtracted, which were obtained along the galaxy's photometric major axis.

   \begin{figure*}
   \centering
   \includegraphics[angle=90,scale=1.0]{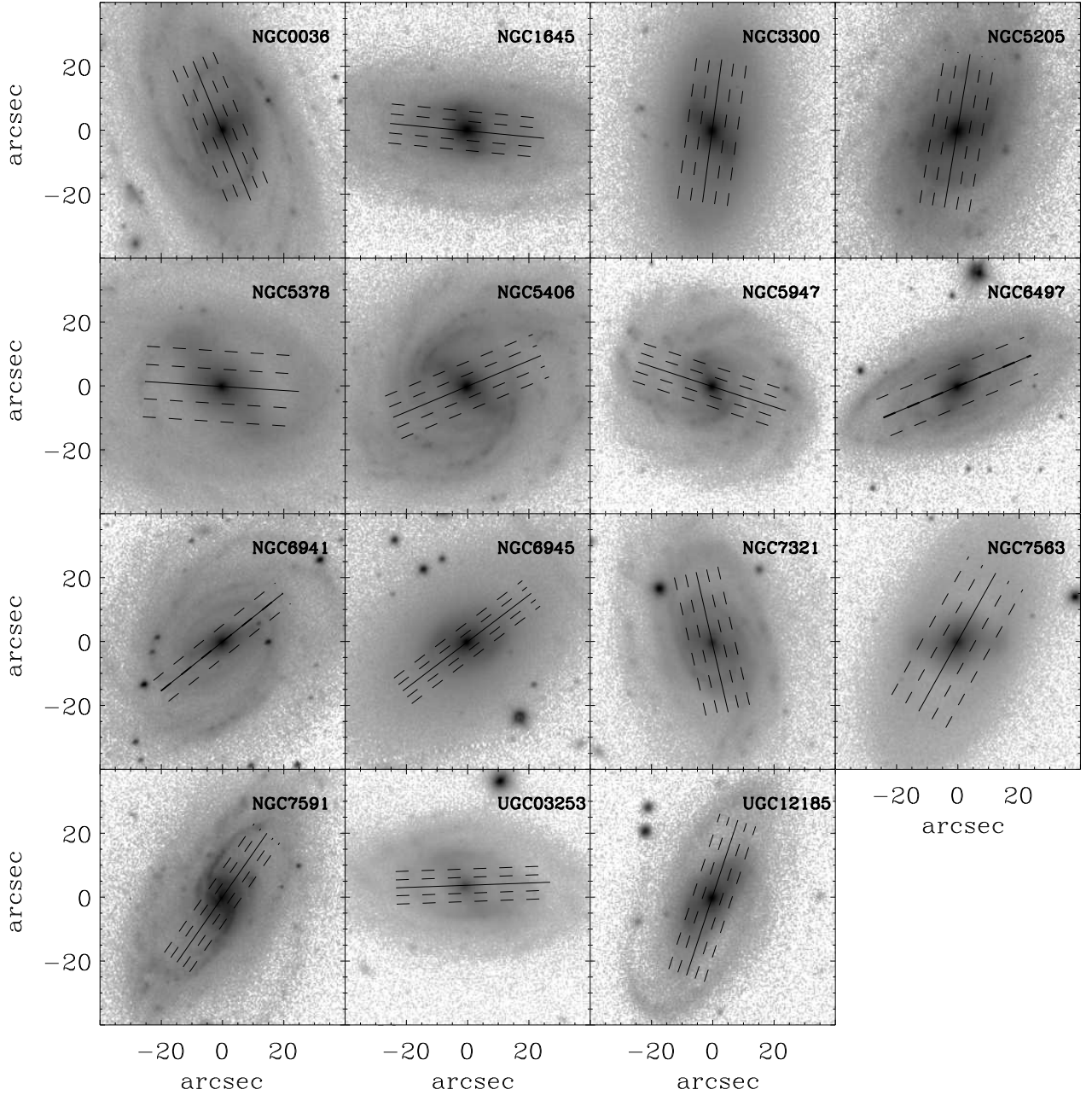}
   \caption{SDSS $r$-band images of the CALIFA  barred galaxies presented in this study. In all images, north is up and east is left. The full and dashed lines represent the considered slits for measuring the kinematic and photometric integrals. The full line also represents the line of nodes of the galaxy.}
              \label{f1}%
    \end{figure*}
%
 
   \begin{figure*}
   \centering
   \includegraphics[angle=0,scale=0.7]{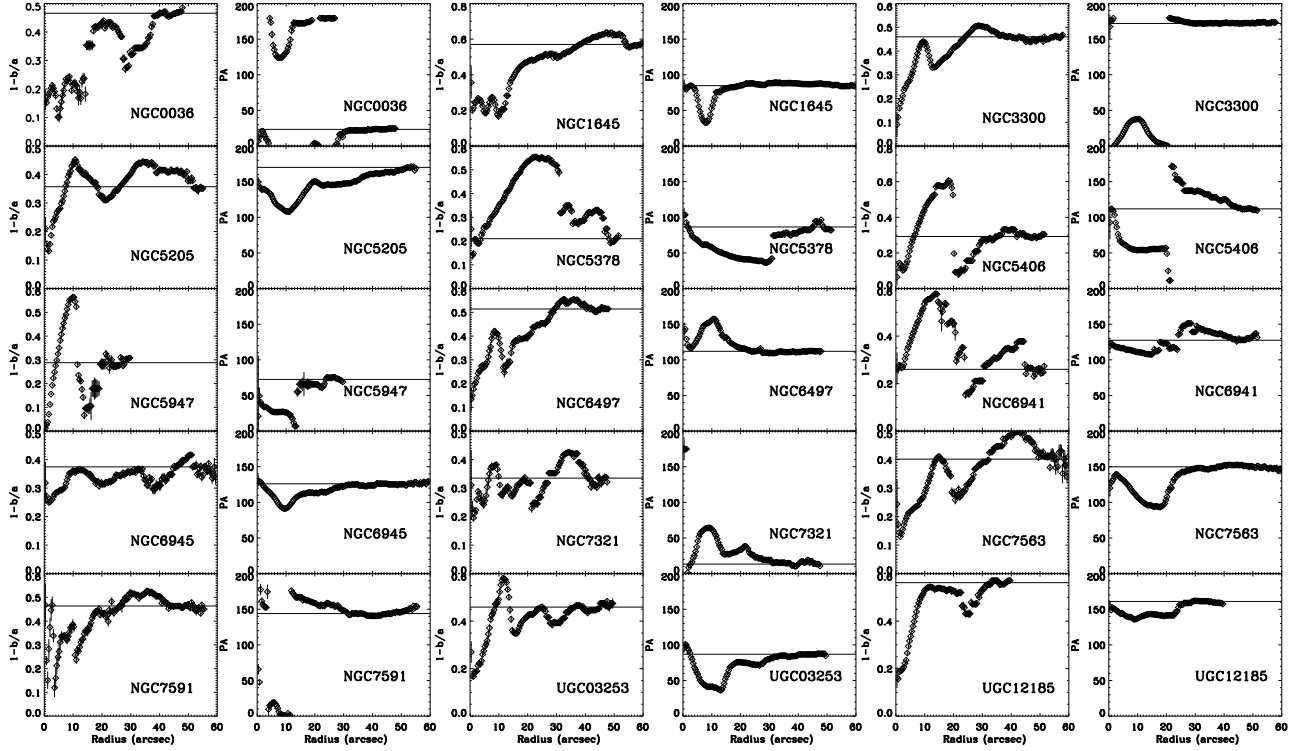}
   \caption{Isophotal ellipticity and PA radial profiles from ellipse-fitting of the barred galaxies of the sample. The horizontal lines show the measured ellipticity and PA corresponding to the disc.}
              \label{f1_phot}%
    \end{figure*}

  \begin{table}
      \caption[]{Bar de-projected radius measurements for the CALIFA barred galaxy sample}
         \label{tb2}
         \centering
         \begin{tabular}{c c c c c }
            \hline
            \hline
           Galaxy &  $a_{\rm b,1}$ & $a_{\rm b,2}$ & $a_{\rm b,3}$  & $a_{\rm b}$  \\
                  & (arcsec)      &   (arcsec)  & (arcsec)      & (arcsec) \\
             (1)     &    (2)        &     (3)     &    (4)        &    (5) \\
            \hline 
            NGC~0036 & 20.8 & 24.7 & 15.1 & 20.2$^{+5.1}_{-4.5}$\\ 
            NGC~1645 &  13.6 & 17.3 & 17.0 & 16.0$^{+2.4}_{-1.3}$\\
            NGC~3300 &  13.4 & 17.6 & 17.4 & 16.1$^{+2.7}_{-1.5}$\\
            NGC~5205 &  14.9 & 18.3 & 19.8 & 17.7$^{+2.8}_{-2.1}$\\
            NGC~5378 &  25.3 & 34.4 & 23.4 & 27.7$^{+4.3}_{-6.7}$\\
            NGC~5406&  20.0 & 23.2 & 20.0 & 21.1$^{+1.1}_{-2.1}$\\
            NGC~5947 &  9.6 & 12.5 & 10.7 & 10.9$^{+1.3}_{-1.6}$\\
            NGC~6497 &  12.6 & 15.6 & 16.0 & 14.7$^{+2.1}_{-1.3}$\\
           NGC~6941 &  14.6 & 18.0 & 15.0 & 15.9$^{+1.3}_{-2.1}$\\ 
            NGC~6945 &  13.8 & 16.4 & 19.2 & 16.5$^{+2.7}_{-2.7}$\\
            NGC~7321 &  10.4 & 14.1 & 11.7 & 12.1$^{+1.7}_{-2.0}$\\
            NGC~7563 &  22.8 & 30.9 & 22.7 & 25.5$^{+2.8}_{-5.4}$\\
            NGC~7591 &  11.0 & 15.0 & 14.5 & 13.5$^{+2.5}_{-1.5}$\\
           UGC~03253 &  14.5 & 18.0 & 14.9 & 15.8$^{+1.3}_{-2.2}$\\
           UGC~12185 &  12.0 & 24.3 & 24.5 & 20.3$^{+8.3}_{-4.2}$\\
            \hline
        \end{tabular}
\tablefoot{Col. (1) Galaxy name. Col. (2): bar radius measured from the maxima of the ellipticity radial profile. Col. (3): bar radius measured when the PA in the bar region changes more than 5$^{\circ}$. Col. (4): bar radius measured from the Fourier bar/interbar intensities. Col. (5): mean of the three bar-length estimations. Errors correspond to the maximum differences between the mean and the three measurements.}
   \end{table}

\section{Results}

\subsection{The bar pattern speed}

The bar pattern speed of the galaxies was measured by applying the TW method as described in Eq. (1). This method requires determining the mean weighted position $(\langle X \rangle)$ and velocity $(\langle V \rangle)$ of stars in several slits oriented along and offset with respect to the LON of the galaxies. We refer to these slits defined in the integral-field datacubes as pseudo slits. Depending on the bar radius and its orientation with the LON, we have used three or five pseudo slits of 1 arcsec width each and a minimum separation of 2 arcsec between them to avoid repeated information. Light and mass were used as two different weights for computing $\langle X \rangle$ and $\langle V \rangle$ for each pseudo slit. 

The high $S/N$ ratio of the broad-band photometric images has led in the past to their being broadly used for computing the photometric integrals of the TW method. Nevertheless, the trace of different stellar populations between the photometric and spectroscopic data, caused by different wavelength coverage and problems with the positioning of the pseudo slits in the photometric images, can affect  the computation of the photometric integrals. These problems can be solved by using integral-field data. In particular, we computed the mean position of stars along the pseudo slits by using the surface brightness distribution obtained directly from the CALIFA datacubes of the galaxies. The surface brightness map of each galaxy was obtained by summing up all the flux from each spectrum of the datacube in a wavelength window of 150 $\AA$ width, and centred at 4575 $\AA$.  

This wavelength window was used because no prominent emission lines are observed in this range.  It is also important to notice that the light distribution does not always trace the mass distribution of the galaxies.  This is especially true in late-type galaxies where star formation is common. To get rid of this, we also computed the mass-weighted mean position of stars by using the profiles extracted along the pseudo-slit positions in the mass maps of the galaxies provided by Gonz\'alez Delgado et al. (2013) and P\'erez et al. (2013). Thus, for each galaxy we have two values of the photometric integrals of the TW: the mass- and light-weighted photometric integrals.

Two different approaches were used to measure the weighted LOS stellar velocity ($\langle V \rangle$) for each pseudo-slit position. The first approach was by computing the integrals in the numerator of Eq. (1) using pseudo slits placed directly in the velocity maps of the galaxies.  The second approach consists in a weighted sum of the raw spectra from the datacube along the pseudo slits. This results in a new, single spectrum for each pseudo slit, which was then analysed using the pPXF method as explained in the previous section. Thus, $\langle V \rangle$ is the radial velocity obtained from the fit to this single spectrum. In the two previous approaches, $\langle V \rangle$ was obtained by also using light and mass as weights. In all cases, Monte Carlo (MC) simulations were used to compute the errors of $\langle V \rangle$.

\begin{figure*}
   \centering
   \includegraphics[angle=90]{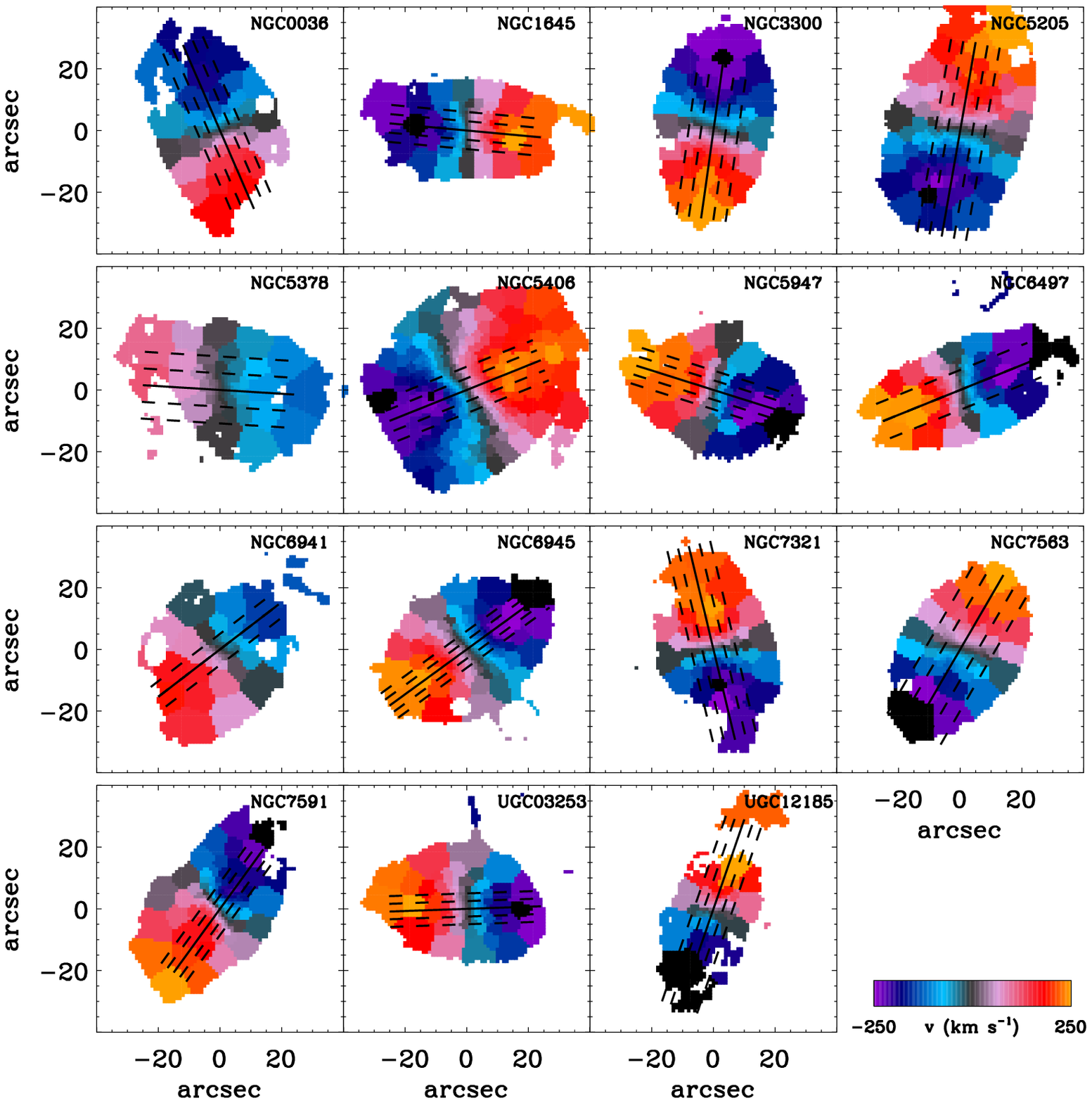}
   \caption{Binned version of the stellar velocity maps of the barred galaxies in our sample. The orientation of the maps is as in Fig. \ref{f1}. Blue colours represent approaching velocities, while red ones show receding velocities. The full and dashed lines represent the considered slits for measuring the kinematic and photometric integrals. The full line also represent the line of nodes of the galaxy.}
              \label{f2}%
    \end{figure*}

Formally, all the TW integrals shown in Eq. (1) are over $-\infty < X < \infty$. But, they can be limited to $-X_{\rm max} < X < X_{\rm max}$ if $X_{\rm max}$ reaches the axisymmetric part of the disc. We used the maximum value of $X_{\rm max}$ allowed by our velocity maps. In most of the cases, this value was $X_{\rm max}=20-30$ arcsec (see Fig. \ref{f2}). For our galaxies we have a mean value of $X_{\rm max}/h_{\rm disc}$=1.6, where $h_{\rm disc}$ is the exponential scale length of the disc of the galaxies. The disc scale length was determined by fitting the outermost regions (outside of the bar region) of  the r-band isophotal surface-brightness profiles of the galaxies by an exponential law \citep[see][]{freeman1970}.

The pattern speed of the galaxies was determined by the slope of the straight line fitted to the $\langle V \rangle$ vs $\langle X \rangle$ points. The uncertainties in $\Omega_{\rm b}$ were obtained by using MC simulations taking the uncertainties in the $PA$ of the galaxies
into account. The $PA$ uncertainties were distributed according to Gaussian distributions. We have four different determinations of the pattern speed ($\Omega_{\rm  b,1}$ to $\Omega_{\rm b,4}$)  of the galaxies depending on both the weight used for computing the TW integrals and the method used for the kinematic integrals. The TW kinematic integrals were obtained by computing the integrals shown in the numerator of Eq. (1), using the velocities provided by the galaxy velocity maps along pseudo-slit positions for $\Omega_{\rm b,1}$ (light-weighted) and $\Omega_{\rm b,3}$ (mass-weighted). The sum of the spectra from the galaxy datacube along the pseudo-slit position was used for computing the kinematic integrals in $\Omega_{\rm b,2}$ (light-weighted) and $\Omega_{\rm b,4}$ (mass-weighted).  Figure \ref{f4} shows the $\langle V \rangle$ vs $\langle X \rangle$ line fits for the sample galaxies.   The linear fits shown in Fig. \ref{f4} do not take the uncertainties of the photometrical integrals into account. The errors of the photometric integrals are in all cases smaller than 0.1 except for one pseudo slit in the case NGC0036, which is 0.14. Table \ref{tb3} shows the values of $\Omega_{\rm b}$ for the CALIFA barred galaxies presented here.

\begin{figure*}
   \centering
   \includegraphics[angle=90]{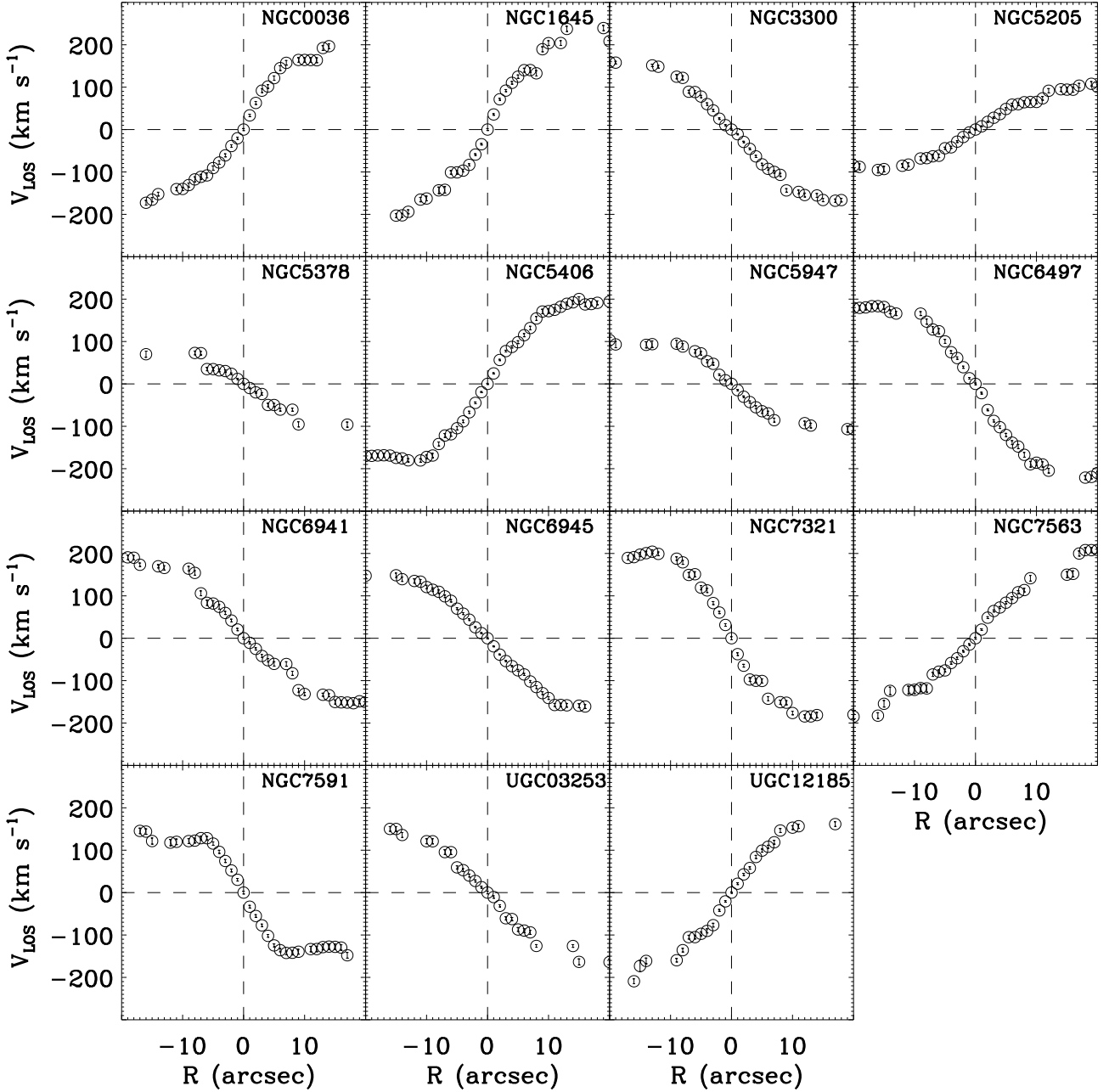}
   \caption{LOS-streaming velocities of the galaxies along their major axis.}
              \label{f3}%
    \end{figure*}

\begin{figure*}
   \centering
   \includegraphics[angle=90]{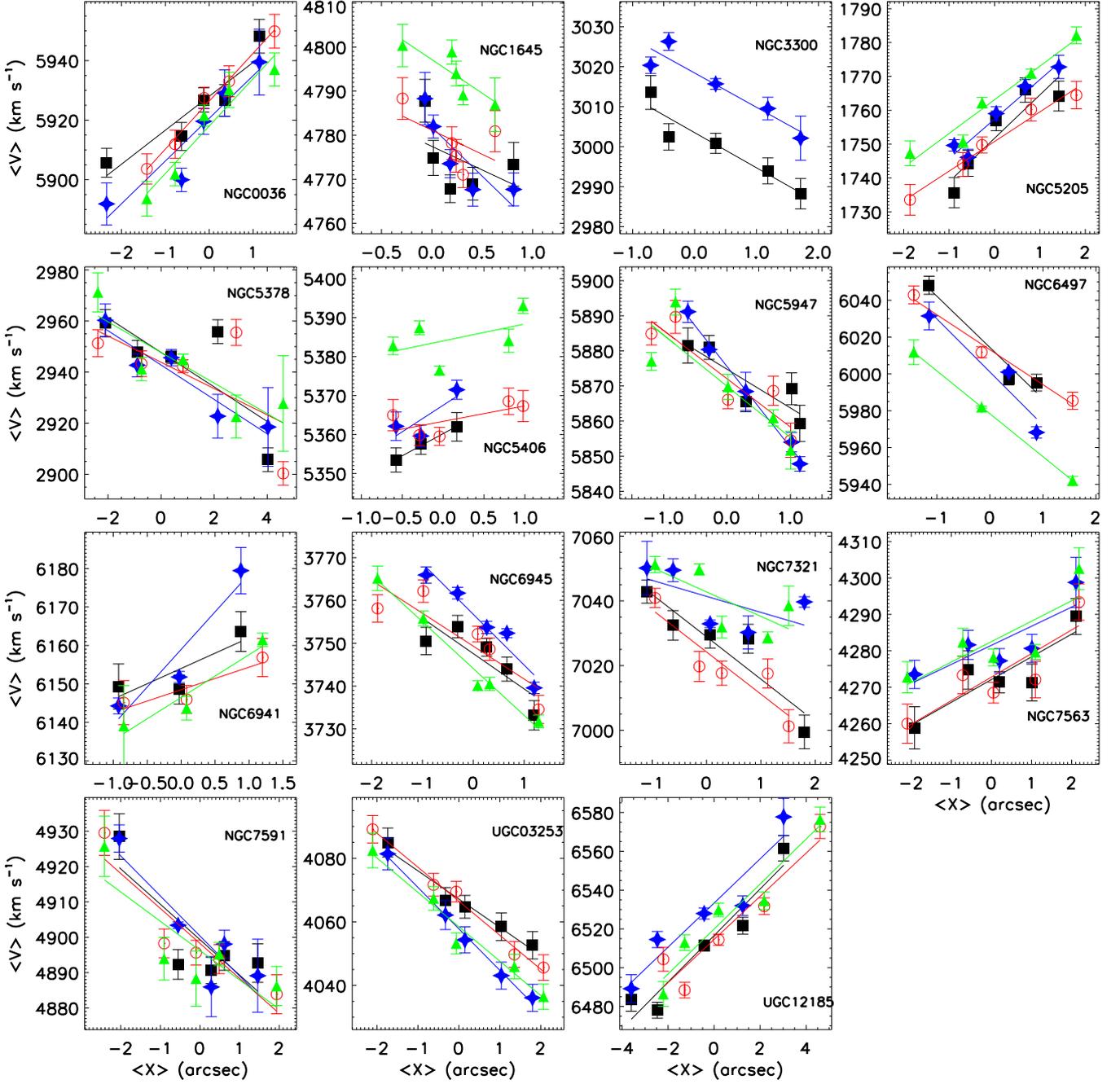}
   \caption{Values of $\langle V \rangle$ vs $\langle X \rangle$ for the galaxies of the sample. The different colours represent the different techniques used for computing the kinematic and photometric integrals of the TW method. Black and green lines represent light-weighted photometric integrals, with kinematic integrals computed using velocity maps or summing spectra along pseudo slits, respectively (corresponding to $\Omega_{\rm b,1}$ and $\Omega_{\rm b,2}$). Red and blue lines represent mass-weighted photometric integrals with kinematic integrals computed using velocity maps or summing spectra along pseudo slits, respectively (corresponding to $\Omega_{\rm b,3}$ and $\Omega_{\rm b,4}$).}
              \label{f4}%
    \end{figure*}

\begin{figure*}
   \centering
   \includegraphics[angle=0,scale=1.0]{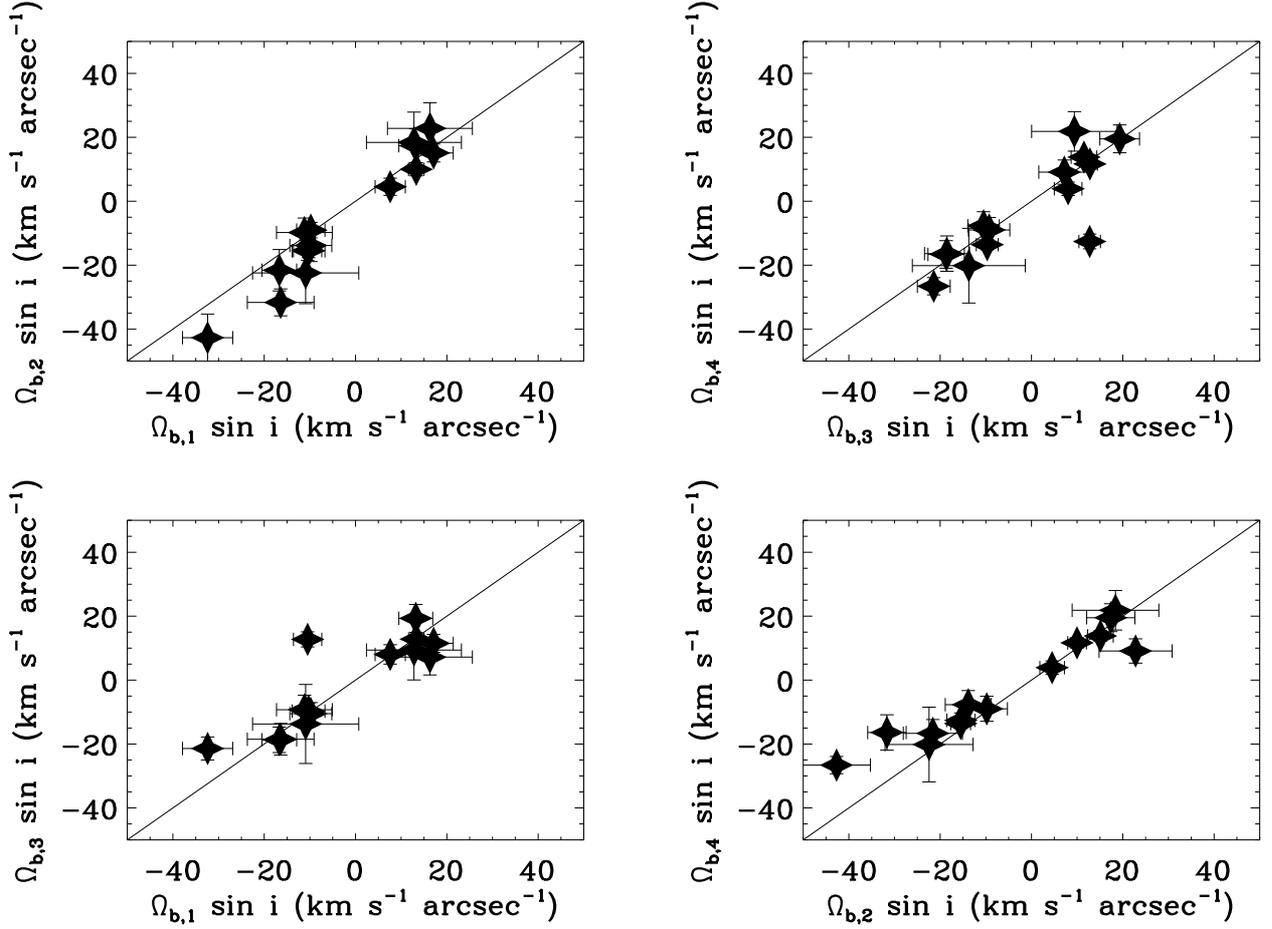}
   \caption{Comparison of the bar pattern speeds of the galaxies using different methods.}
              \label{fomega}%
    \end{figure*}

Figure \ref{fomega} shows the comparison of the values of $\Omega_{\rm b} \sin i$ obtained by different methods. In general, no systematic and/or significant differences in the value of  $\Omega_{\rm b} \sin i$ can be seen. Only three galaxies (NGC5406, NGC5947, and NGC6497) present some value of $\Omega_{b}$ that does not agree, within the errors, with the others. It is not clear to us which is the main reason for this difference, since these galaxies do not present irregular features in their rotation velocity maps (see Fig. \ref{f2}) and/or velocity curves (see Fig. \ref{f3}). In addition, they are galaxies of different morphological types, which seems to indicate that these differences are not related to different amounts of dust, gas, or star formation in their discs. These differences could simply reflect the observational uncertainties in the determination of $\Omega_{b}$.

%

%

%

%

  \begin{table*}
      \caption[]{Pattern speed and corotation radius of the CALIFA barred galaxies}
         \label{tb3}
         \centering
\small
         \begin{tabular}{c c c c c c c c c}
            \hline
            \hline
           Galaxy &  $\Omega_{\rm b,1}$ & $\cal{R}_{\rm CR,1}$ &$\Omega_{\rm b,2}$ & $\cal{R}_{\rm CR,2}$& $\Omega_{\rm b,3}$  & $\cal{R}_{\rm CR,3}$& $\Omega_{\rm b,4}$  & $\cal{R}_{\rm CR,4}$\\
                     &   (km s$^{-1}$ arcsec$^{-1}$) & (arcsec) &   (km s$^{-1}$ arcsec$^{-1}$) &   (arcsec) & (km s$^{-1}$ arcsec$^{-1}$) & (arcsec) & (km s$^{-1}$ arcsec$^{-1}$) & (arcsec) \\
            \hline 
            NGC~0036 &  13.2$\pm$3.7  & 16.9$^{+6.6}_{-4.8}$  &  17.4$\pm$5.2 & 12.6$^{+5.4}_{-3.9}$ &  19.3$\pm$4.3  & 14.2$^{+3.2}_{-2.6}$  & 19.5$\pm$4.3   & 11.1$^{+3.9}_{-3.0}$ \\ 
            NGC~1645 & -10.9$\pm$11.6 & 18.9$^{15.5}_{7.6}$   & -22.4$\pm$9.4 & 12.1$^{6.7}_{3.8}$   & -13.7$\pm$12.4 & 19.1$^{11.5}_{5.3}$  & -20.2$\pm$11.7 & 13.1$^{10.0}_{4.8}$\\
            NGC~3300 &  -9.8$\pm$3.1  & 23.2$^{+8.9}_{-5.9}$  &  -9.0$\pm$2.4 & 24.5$^{8.2}_{6.1}$   &                &                     &                & \\
            NGC~5205 &  17.1$\pm$4.3  & 10.2$^{+3.2}_{2.4}$   &  15.1$\pm$2.8 & 11.3$^{3.0}_{2.5}$   &  11.5$\pm$2.8  & 11.2$^{8.4}_{4.7}$    & 13.8$\pm$1.7  & 12.3$^{2.7}_{2.4}$ \\
            NGC~5378 & -11.2$\pm$6.1  & 16.3$^{+13.0}_{-6.6}$ &  -9.8$\pm$4.5 & 18.0$^{+12.0}_{-7.3}$ &  -9.3$\pm$4.5  & 17.5$^{+15.0}_{-8.7}$ & -9.0$\pm$3.9  & 19.7$^{+12.4}_{-7.8}$ \\
            NGC~5406 &  16.3$\pm$9.3  & 15.5$^{+12.0}_{-5.3}$ &  22.8$\pm$8.0 & 11.0$^{+4.8}_{-3.0}$ &    7.2$\pm$5.6  & 14.7$^{+36.3}_{8.7}$  & 9.1$\pm$ 3.8  & 27.4$^{+14.7}_{-8.3}$ \\
            NGC~5947 & -16.4$\pm$7.3  & 11.6$^{+8.3}_{-4.8}$  & -31.7$\pm$4.2 & 5.8$^{+2.4}_{-2.3}$  &  -18.5$\pm$4.9  & 10.7$^{+4.4}_{-3.6}$  & -16.4$\pm$5.6 & 11.2$^{+6.8}_{-4.8}$ \\
            NGC~6497 & -32.4$\pm$5.5  &  7.4$^{+2.6}_{-2.0}$  & -42.7$\pm$7.4 & 5.5$^{+2.1}_{-1.6}$  &  -21.4$\pm$3.6  & 8.7$^{+6.3}_{-4.1}$   & -26.6$\pm$2.7 & 8.8$^{+3.0}_{-2.4}$ \\
           NGC~6941  &  12.8$\pm$10.4 & 14.6$^{+14.4}_{-5.8}$ &  18.4$\pm$9.5 & 10.7$^{+7.4}_{3.7}$  &    9.4$\pm$9.3  & 14.1$^{+19.9}_{-6.8}$ & 21.9$\pm$6.2  & 9.2$^{+3.1}_{-2.2}$ \\ 
            NGC~6945 & -10.3$\pm$3.6  & 20.2$^{+8.9}_{-5.7}$  & -15.5$\pm$2.1 & 13.0$^{+2.9}_{-2.5}$  &  -9.7$\pm$2.4  & 20.3$^{+7.2}_{-5.4}$  & -13.6$\pm$1.6 & 14.7$^{+3.0}_{-2.6}$ \\
            NGC~7321 & -16.7$\pm$3.8  & 15.6$^{+5.7}_{-4.4}$  & -21.6$\pm$6.5 & 11.8$^{+5.4}_{-4.0}$  & -18.7$\pm$4.0  & 14.6$^{+3.9}_{-3.3}$  & -16.7$\pm$4.3 & 15.3$^{+6.2}_{-4.9}$ \\
            NGC~7563 &   7.6$\pm$3.3  & 33.2$^{+18.4}_{-10.2}$&   4.5$\pm$2.7 & 45.4$^{+19.6}_{-15.8}$ &   8.0$\pm$3.0 & 32.2$^{+13.5}_{-8.4}$  & 3.9$\pm$2.1   & 51.3$^{+16.0}_{-17.1}$ \\
            NGC~7591 &  -9.8$\pm$4.6  & 18.5$^{+10.7}_{-6.2}$  & -13.9$\pm$5.0 & 13.2$^{+6.6}_{-4.3}$  &  -10.5$\pm$3.5 & 18.3$^{+7.0}_{-4.9}$  & -7.7$\pm$4.4  & 20.6$^{+10.6}_{-7.4}$\\
           UGC~03253 & -10.5$\pm$3.1  & 18.0$^{+6.6}_{-4.5}$  & -15.5$\pm$3.1 & 11.9$^{+3.2}_{-2.7}$  &  -12.8$\pm$2.4 & 16.2$^{+2.3}_{-2.0}$  & -12.6$\pm$2.3 & 14.6$^{+3.8}_{-3.2}$ \\
           UGC~12185 &  13.3$\pm$1.7  & 16.7$^{+8.6}_{-6.4}$  &  10.0$\pm$2.0 & 21.6$^{12.3}_{9.1}$   &   12.8$\pm$1.6 & 16.6$^{+9.2}_{-6.9}$  & 11.6$\pm$1.7  & 18.4$^{+10.1}_{-7.6}$ \\
            \hline
        \end{tabular}
\begin{tablenotes}
\small
\item Note: NGC 3300 has no values of $\Omega_{b,3}$ and $\Omega_{b,4}$ because the mass distribution of this galaxy is not available.
\end{tablenotes}
   \end{table*}

\subsection{Determination of ${\cal R}=R_{\rm CR}/a_{\rm b}$}
 
The dimensionless quantity ${\cal R}=R_{\rm CR}/a_{\rm b}$ is defined as the ratio between the corotation radius ($R_{\rm CR}$) of the galaxy and the bar radius ($a_{\rm b}$). Unlike the bar pattern speed, measure ${\cal R}$ requires some modelling to recover the rotation curve from the observed stellar streaming velocities.

The corotation radius is derived from the circular velocity of the galaxy. To obtain the circular velocity ($V_{\rm c}$) from the observed stellar streaming velocity ($V_{*}$), the asymmetric drift correction is needed. One of the aims of the present work has been to compare our pattern speeds for SBb and SBbc galaxies with those from early-type systems from the literature.To keep the homogeneity in this comparison, we followed the same approximation for the asymmetric drift correction as in previous works \citep[see][]{debattista2002, aguerri2003, corsini2003}. Thus, the asymmetric drift equation becomes

\begin{equation}
V_{\rm c}^{2}=V_{*}^{2} + \frac{\sigma_{\rm obs}^{2}}{\sin^{2} i (1+2\alpha^{2} \cot^{2}i)} \left[2R \left(\frac{1}{R_{d}}+\frac{2}{R_{\sigma}}\right)-1\right]
\end{equation}

where $R$, $V_{*}$, and $\sigma_{obs}$ are the distance to the galaxy centre, the LOS velocity, and velocity dispersion along the major axis of the galaxy, respectively. Parameter $\alpha$ is defined as the ratio between the perpendicular and radial disc velocity dispersions. The value of $\alpha$ can vary from early- to late-type galaxies. We have used the values given by Gerssen \& Shapiro (2012) for the different morphological types of galaxies. In particular, we have used $\alpha=0.85\pm0.15$ for SB0-SB0/a, $\alpha=0.86\pm0.24$ for SBa-SBab, and  $\alpha=0.62\pm0.2$ for SBb-SBbc. In addition, $R_{d}$ and $R_{\sigma}$ are the scale lengths of the surface brightness and velocity dispersion profiles, assuming an exponential law for the disc \cite[e.g.,][]{freeman1970}.

\begin{figure}
   \centering
   \includegraphics[angle=0,scale=0.5]{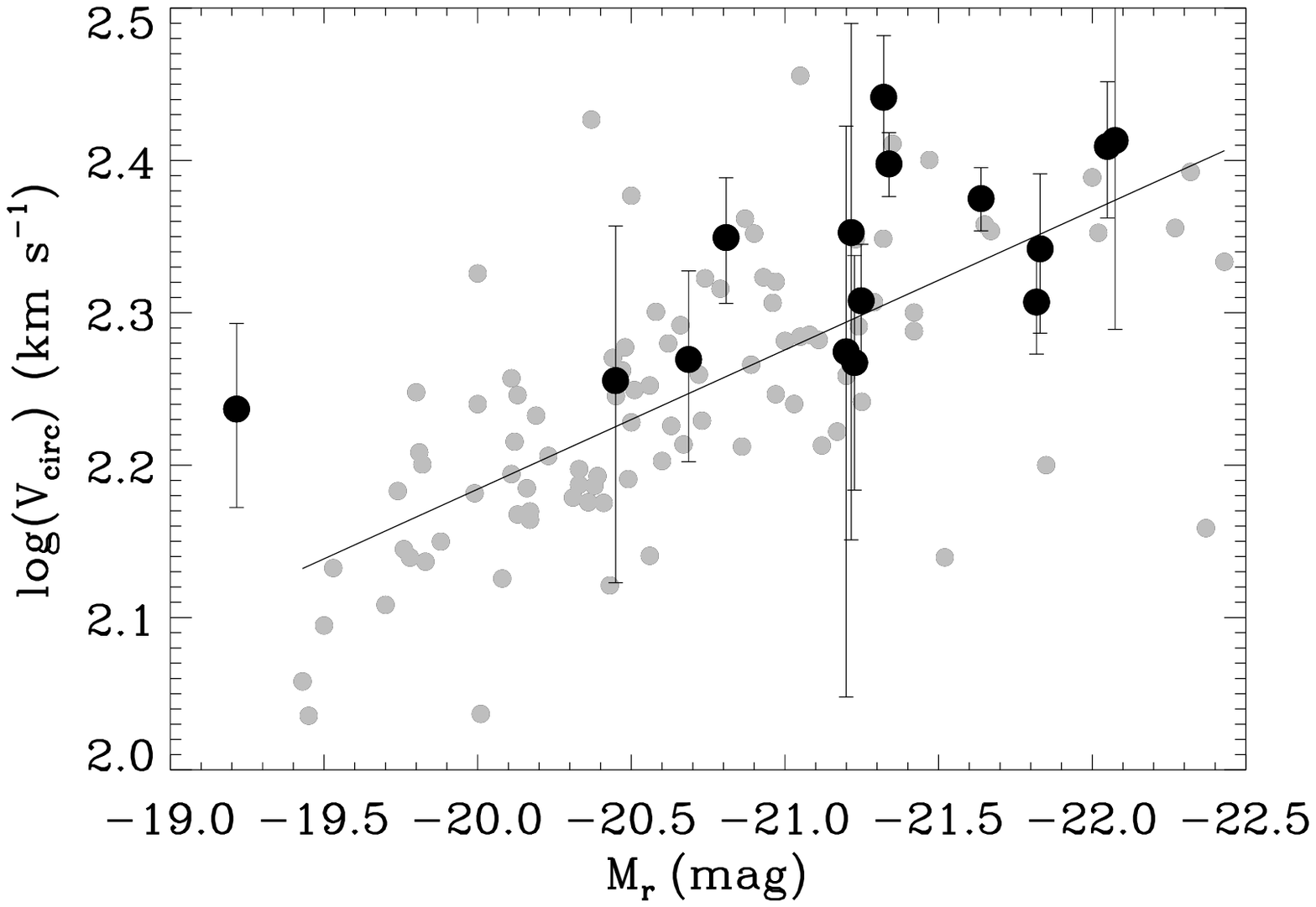}
   \caption{Tully-Fisher relation of our CALIFA galaxies (black symbols) and spiral galaxies  from Reyes et al. (2011) (grey symbols). The circular velocity of the CALIFA galaxies was optained from the asymmetric drift correction applied in this study. The full line represent the fit of the TF relation for the galaxies from Reyes et al. (2011).}
              \label{tf}%
    \end{figure}

This correction was applied to the velocity data along the major axis of the galaxy discs and beyond to the bar region. The value of $V_{\rm c,flat}$ for the galaxies was obtained by averaging $V_{\rm c}$ outside of the bar region. The Tully-Fisher relation \citep[TF; ][]{tully1977} shows that there is a correlation between the circular velocity of the galaxies at large radii and their absolute magnitudes. We located our galaxies in the TF relation by using the $V_{\rm c,flat}$ obtained from the asymmetric drift correction. Figure \ref{tf} shows the TF relation for our galaxies and other spiral galaxies obtained from the literature \citep[see][]{reyes2011}. We notice that our galaxies are located within the noise of the TF  relation. This indicates that the $V_{\rm c,flat}$ we obtained is consistent with what is proposed by the TF relation for spiral galaxies.

In addition, we found in the literature the maximum circular velocity measured from HI data for five of our galaxies \citep[see][]{theureau1998}. In all cases, this HI rotation velocity agrees within the errors with our values of $V_{c,flat}$. For these five galaxies, the ratios between the maximum circular HI rotation velocity ($V_{c,HI}$) and our $V_{c,flat}$ are $V_{c, HI}/V_{c,flat}=1.12, 1.14, 0.99, 0.88$, and $1.05$ for NGC0036, NGC6941, NGC7321, NGC7563, and UGC03253. This indicates a maximun uncertainty of $14\%$ in our flat rotation determinations.

The value of the corotation radius is then given by $R_{\rm CR}=V_{\rm c,flat}/\Omega_{\rm b}$. This assumes that the rotation curve of the galaxies is flat. The values of $R_{\rm CR}$ for our galaxies are given in Table \ref{tb3}. We finally measure ${\cal R}$ as $R_{\rm CR}/a_{\rm b}$. The values of ${\cal R}$ are given in Table \ref{tb4}. Errors in Tables \ref{tb3} and \ref{tb4} were  computed by Monte-Carlo simulations.

%

  \begin{table}
      \caption[]{Dimensionless quantity ${\cal R}$ for the CALIFA barred galaxies}
         \label{tb4}
         \centering
\small
         \begin{tabular}{c c c c c}
            \hline
            \hline
           Galaxy &  $\cal{R}$$_{1}$ & $\cal{R}$$_{2}$ & $\cal{R}$$_{3}$ & $\cal{R}$$_{4}$\\
            \hline 
            NGC~0036  & 0.9$^{+0.3}_{-0.2}$ & 0.6$^{+0.3}_{-0.2}$ & 0.6$^{+0.2}_{-0.1}$ & 0.6$^{+0.2}_{-0.1}$ \\ 
            NGC~1645  & 1.3$^{+1.7}_{-0.5}$ & 0.8$^{+0.4}_{-0.2}$ & 1.1$^{+1.1}_{-0.4}$ & 0.8$^{+0.7}_{-0.3}$ \\
            NGC~3300  & 1.5$^{+0.6}_{-0.4}$ & 1.6$^{+0.5}_{-0.4}$ & & \\
            NGC~5205  & 0.6$^{+0.2}_{-0.1}$ & 0.7$^{+0.2}_{-0.1}$ & 0.9$^{+0.3}_{-0.2}$ & 0.7$^{+0.1}_{-0.1}$\\
            NGC~5378  & 0.6$^{+0.5}_{-0.2}$ & 0.6$^{+0.4}_{-0.2}$ & 0.6$^{+0.5}_{-0.3}$ & 0.7$^{+0.4}_{-0.3}$ \\
            NGC~5406  & 0.7$^{+0.6}_{-0.2}$ & 0.5$^{+0.2}_{-0.1}$ & 1.4$^{+1.5}_{-0.6}$ & 1.3$^{+0.7}_{-0.4}$ \\
            NGC~5947  & 1.0$^{+0.8}_{-0.4}$ & 0.5$^{+0.2}_{-0.2}$ & 0.9$^{+0.5}_{-0.4}$ & 1.0$^{+0.6}_{-0.4}$ \\
            NGC~6497  & 0.4$^{+0.1}_{-0.1}$ & 0.3$^{+0.1}_{-0.1}$ & 0.6$^{+0.2}_{-0.1}$ & 0.5$^{+0.1}_{-0.1}$ \\
            NGC~6941  & 0.9$^{+0.9}_{-0.3}$ & 0.6$^{+0.5}_{-0.2}$ & 1.1$^{+1.3}_{-0.4}$ & 0.6$^{+0.2}_{-0.1}$ \\ 
            NGC~6945  & 1.2$^{+0.6}_{-0.3}$ & 0.8$^{+0.2}_{-0.1}$ & 1.3$^{+0.4}_{-0.3}$ & 0.9$^{+0.2}_{-0.1}$ \\
            NGC~7321  & 1.3$^{+0.5}_{-0.3}$ & 1.0$^{+0.4}_{-0.3}$ & 1.1$^{+0.4}_{-0.3}$ & 1.2$^{+0.5}_{-0.4}$ \\
            NGC~7563  & 1.2$^{+0.7}_{-0.3}$ & 1.9$^{+1.7}_{-0.7}$ & 1.1$^{+0.6}_{-0.3}$ & 2.3$^{+1.8}_{-0.8}$ \\
            NGC~7591  & 1.4$^{+1.0}_{-0.5}$ & 1.0$^{+0.5}_{-0.3}$ & 1.3$^{+0.6}_{-0.4}$ & 1.7$^{+1.3}_{-0.6}$ \\
           UGC~03253  & 1.1$^{+0.4}_{-0.3}$ & 0.7$^{+0.2}_{-0.2}$ & 0.9$^{+0.2}_{-0.2}$ & 0.9$^{+0.2}_{-0.2}$ \\
           UGC~12185  & 0.9$^{+0.4}_{-0.3}$ & 1.2$^{+0.6}_{-0.5}$ & 0.9$^{+0.4}_{-0.4}$ & 1.0$^{+0.5}_{-0.4}$ \\
            \hline
        \end{tabular}
   \end{table}

\subsection{Fast or slow bars}
The galaxy sample presented in this work complemented by those available in the literature (see Table 1 from Corsini 2011) make a total of 32 galaxies with bar pattern speed measured by the TW method. This represents an appropriate sample for inferring strong observational constraints on the value of ${\cal R}$ in bright barred galaxies. Table \ref{tb4} shows that all bars have values of ${\cal R}$, within the errors, compatible with being fast bars ($R<1.4$). We computed the probability of having $R>1.4$ in the whole sample by using Monte Carlo simulations that take the quoted errors into account. We cannot rule out a fast bar in any of our galaxies (at 95\% probability).  

We have also determined the mean value of ${\cal R}$ for the total sample of 32 galaxies with $\Omega_{\rm b}$ determined by the TW method. Taking our four different measurements of ${\cal R}$
into account, we obtained $<{\cal R}_{1}>=1.2^{+0.7}_{-0.5}$, $<{\cal R}_{2}>=1.0^{+0.7}_{-0.4}$, $<{\cal R}_{3}>=1.2^{+0.7}_{-0.4}$, and $<{\cal R}_{4}>=1.1^{+0.7}_{-0.4}$. Table 4 shows that some of our values of ${\cal R}$ have large errors. We selected a subsample of galaxies with small uncertainties in ${\cal R}$. In particular, we considered those galaxies from the literature and from our measurements with uncertainties smaller than 30$\%$ in determining ${\cal R}$. For this subsample of galaxies, we have  $<{\cal R}_{1}>=1.0^{+0.3}_{-0.3}$, $<{\cal R}_{2}>=0.9^{+0.3}_{-0.2}$, $<{\cal R}_{3}>=1.0^{+0.3}_{-0.2}$, and $<{\cal R}_{4}>=0.9^{+0.3}_{-0.2}$. In all cases, the mean values indicate that bars finish near corotation in agreement with numerical simulations of barred galaxies \citep[see][]{athanassoula1992}. 

\subsection{Dependence of the bar pattern speed on the morphological type of the galaxy}

\begin{figure*}
   \centering
   \includegraphics[angle=0]{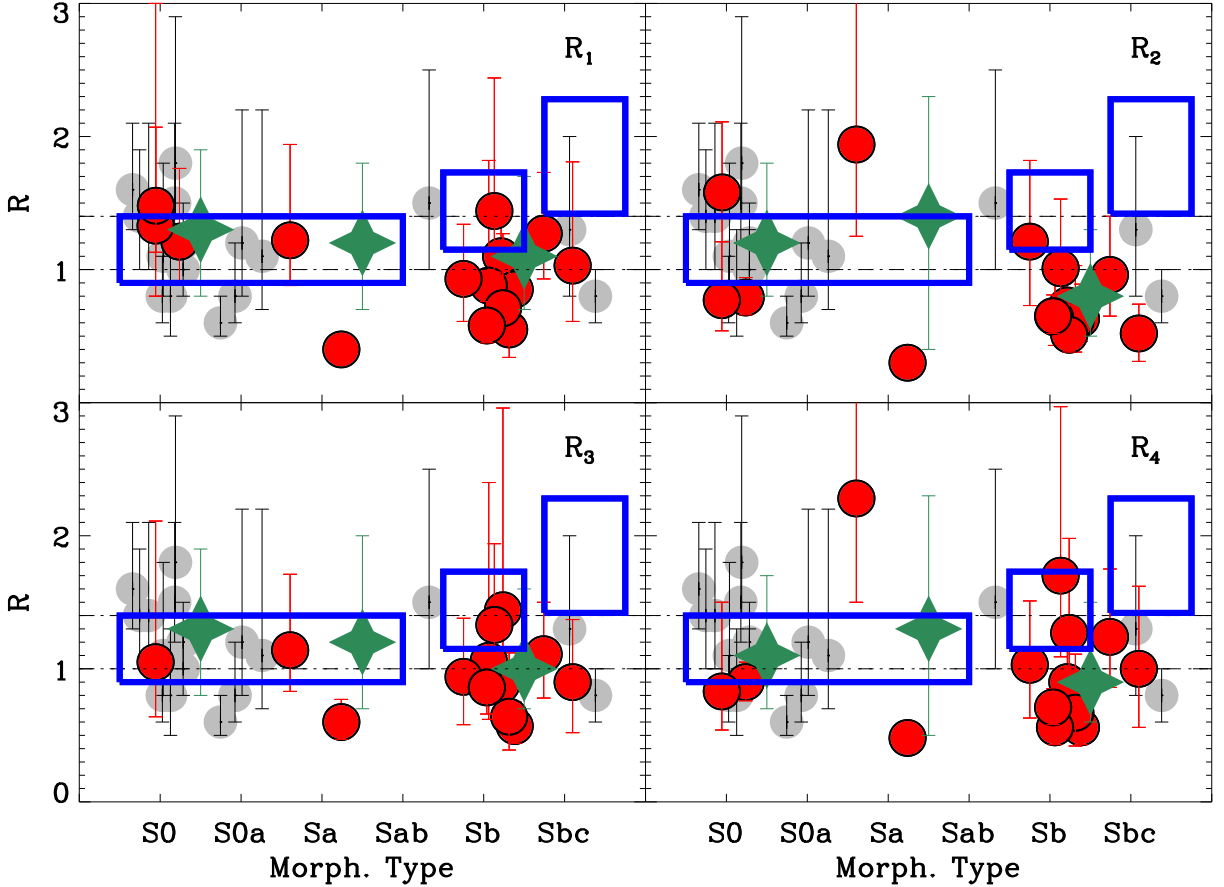}
   \caption{${\cal R}$ value as a function of the Hubble type for all galaxies with pattern speed measured by the TW method. The grey points represent galaxies from the literature compiled by Corsini (2011), which excludes the sample of Rautiainen et al. (2008), which we indicate by the blue rectangles. The red symbols represent the ${\cal R}$ values of our CALIFA sample. The green stars represent the mean values of ${\cal R}$ for SB0-SB0/a, SBa-SBab, and SBb-SBbc galaxies with $\Omega_{b}$ determined by the TW method. The blue rectangles represent the mean and dispersion (width of the rectangle) values of ${\cal R}$ from Rautiainen et al. (2008). For clarity reasons, ${\cal R}$ values have been slightly shifted with respect to its Hubble type.} 
              \label{r_morpho}%
    \end{figure*}

Figure \ref{r_morpho} shows the variation in ${\cal R}$ with the Hubble morphological type for the 32 galaxies with $\Omega_{\rm b}$ determined by the TW method.  We also computed the mean values of ${\cal R}$ for galaxies in three Hubble-type bins SB0-SB0/a, SBa-SBab, and SBb-SBbc. Table \ref{tb5} shows these mean values. The mean values of ${\cal R}$ for SBb-SBbc galaxies are always lower than for early-type ones. Nevertheless, this trend is not significant when considering the errors.  

As previously noted, the errors showed in Table \ref{tb5} are large. We investigated whether these large errors in ${\cal R}$ are masking a trend with Hubble type. To analyse this possibility, we have also studied the trend in ${\cal R}$ with the Hubble type for the subsample of galaxies with uncertainties smaller than $30\%$ in the determining ${\cal R}$. In this case, we have only compared SB0-SB0/a and SBb-SBbc Hubble types, because we have no galaxies in the morphological bin SBa-SBab. Table \ref{tb6} shows these values. In this case, the errors are considerably smaller. Although the mean values of ${\cal R}$ are also systematically smaller for late-type galaxies, these differences are again compatible within the uncertainties. We can therefore conclude that no significant trend is observed in the ${\cal R}$ values of early and late-type galaxies. 

For the sake of comparison in Fig. \ref{r_morpho}, we have also included  the mean values of ${\cal R}$ in three different morphological ranges (SB0+SB0/a, SBa+SBab, and SBb+SBbc) obtained through hydrodynamical simulations by Rautiainen et al. (2008). In contrast to the results shown by Rautiainen et al. (2008), the application of the TW method does not result in any late-type galaxy with slow bars as predicted by Rautiainen et al. (2008).

  \begin{table}
      \caption[]{Mean values of ${\cal R}$ for different Hubble types.}
         \label{tb5}
         \centering
\small
         \begin{tabular}{c c c c c c}
            \hline
            \hline
           Hubble type &  ${\cal R}_{1}$ & ${\cal R}_{2}$ & ${\cal R}_{3}$ & ${\cal R}_{4}$ & $N_{gal}$\\
            \hline 
                SB0-SB0/a & $1.3^{+0.6}_{-0.5}$ & $1.2^{+0.6}_{-0.4}$ & $1.3^{+0.6}_{-0.5}$ & $1.2^{+0.6}_{-0.4}$ & 17\\
                SBa-SBab & $1.2^{+0.6}_{-0.5}$ & $1.4^{+0.9}_{-1.0}$ & $1.2^{+0.8}_{-0.5}$ & $1.3^{+1.0}_{-0.8}$ & 3\\
                SBb-SBbc & $1.1^{+0.6}_{-0.4}$ & $0.8^{+0.5}_{-0.3}$ & $1.0^{+0.6}_{-0.3}$ & $0.9^{+0.6}_{-0.3}$ & 12 \\
                All types & $1.2^{+0.7}_{-0.5}$ & $1.0^{+0.7}_{-0.4}$ & $1.2^{+0.7}_{-0.4}$ & $1.1^{+0.7}_{-0.4}$ & 32 \\
            \hline
        \end{tabular}
\begin{tablenotes}
\small
\item Note: $N_{gal}$ represents the number of galaxies for each morphological type bin.
\end{tablenotes}
   \end{table}

  \begin{table}
      \caption[]{Mean values of ${\cal R}$ for different Hubble types for the subsample of galaxies with smaller uncertainties in the determination of ${\cal R}$.}
         \label{tb6}
         \centering
\small
         \begin{tabular}{c c c c c c}
            \hline
            \hline
           Hubble type &  ${\cal R}_{1}$ & ${\cal R}_{2}$ & ${\cal R}_{3}$ & ${\cal R}_{4}$ & $N_{gal}$\\
            \hline 
                SB0-SB0/a & $1.2^{+0.2}_{-0.2}$ & $1.0^{+0.3}_{-0.2}$ & $1.2^{+0.2}_{-0.2}$ & $1.1^{+0.3}_{-0.2}$ & 13 \\
                SBb-SBbc & $0.9^{+0.3}_{-0.2}$ & $0.8^{+0.2}_{-0.1}$ & $0.9^{+0.2}_{-0.1}$ & $0.8^{+0.2}_{-0.1}$  & 10\\
                All types & $1.0^{+0.3}_{-0.3}$ & $0.9^{+0.3}_{-0.2}$ & $1.0^{+0.3}_{-0.2}$ & $0.9^{+0.3}_{-0.2}$ & 23\\
            \hline
        \end{tabular}
   \end{table}

\section{Discussion}

\subsection{Uncertainties in determiningf the bar pattern speed.}

The pattern-speed measurements using the TW method are sensitive to the determination of the PA of the galaxies. Debattista (2002) tested this dependence using N-body numerical simulations. He found that, for PA errors of about 5$^{\circ}$, the scatter in ${\cal R}$ is $\sim$0.44. The PA of our galaxies was measured by averaging the PA of the outermost isophotes fitted on SDSS $r$-band images.  The CALIFA values of $i$ and PA are consistent within the errors with those reported in the literature. In particular, the median value of the absolute difference between the inclination of the galaxies reported by LEDA \footnote{We acknowledge the usage of the HyperLeda database (http://leda.univ-lyon1.fr)} \citep[][]{paturel2003} and our values is $\Delta i=|i_{LEDA}-i_{CALIFA}|=2.6^{\circ}$. There is a galaxy (NGC~6945) with a large difference between PA$_{CALIFA}$ and PA$_{LEDA}$. For this galaxy  the value of the PA reported by LEDA is $32^{\circ}\pm66^{\circ}$, which is 94$^{\circ}$ different  from our value. A simple visual inspection of the $r$-band galaxy image showed in Fig. \ref{f1} ruled out the LEDA PA value. The median value is, excluding NGC~6945, $\Delta PA=|PA_{LEDA}-PA_{CALIFA}|=4.0^{\circ}$. 

Late-type galaxies have strong structures in the outer regions of the disc (spiral arms, rings, etc.). These structures can affect the orientation of the outer isophotes and subsequently the PA value of the galaxy. Thus, it could be that the real PA of the discs of our galaxies could be different (more than 3$\sigma$) to what is obtained from averaging the outermost isophotes. We checked this possibility by comparing our photometric PA with the one obtained from the symmetrisation of the stellar velocity maps (see for more details Falc\'on-Barroso et al., 2014, in preparation). The photometric and kinematic PA of the galaxies agree within the 3$\sigma$ errors for all cases where the velocity map extends to the outermost regions of the galaxies where we computed their PAs. 

One of the advantages of integral-field spectroscopy over standard long-slit is that the uncertainties in the PA of the galaxies can be taken into account by computing $\Omega_{\rm b}$ along different directions according to the errors in PA. This is what we did in the present work. In particular, we derived $\Omega_{\rm b}$ in a Monte Carlo fashion by varying the PA in the range PA$\pm \Delta$PA, where $\Delta$PA is the uncertainty of the PA given in Table \ref{tb1}. The value of $\Omega_{\rm b} \sin i$ reported in Table \ref{tb3} corresponds to the mean and dispersion of the values obtained for the different PA values. 

\begin{figure}
   \centering
   \includegraphics[angle=0,scale=0.52]{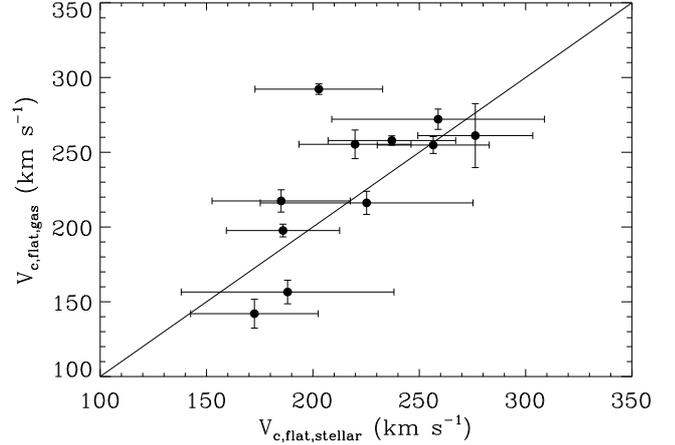}
   \caption{Comparison of $V_{c_flat}$ obtained from stellar and gas kinematics for 11 galaxies of our sample}
              \label{vc_star_gas}%
    \end{figure}

Another uncertainty in determining  $\cal{R}$  could be related to the determination of $V_{c,flat}$, which was obtained after a correction for asymmetric drift to the observed stellar velocities. This is needed due to the stars in the disc of the galaxies do not follow circular velocities. The asymmetric drift correction applied in Sect. 6.2 assumes that the galaxies have thin stellar discs. This approximation could be unreliable for some of the galaxies. In contrast to stars, the gas is almost in circular orbits in galaxy discs. We have the gas velocity curve along the major axis for a total of 11 of our galaxies (see details in Garc\'{\i}a-Lorenzo et al. (2015). Figure \ref{vc_star_gas} shows the comparison between $V_{c,flat}$ obtained from gas and stellar kinematics. For all galaxies except one, NGC6941, the agreement between the two quantities is within the 1$\sigma$ error bars. This galaxy shows $V_{c,flat,gas}/V_{c,flat,stellar} \approx 1.4$. The values of $\cal{R}$ for this galaxy would be in the range $0.8 - 1.5$ if we use $V_{c,flat}$ from gas kinematics.

The convergence of the kinematic integrals could also be another source of uncertainty in the determination of the pattern speed. This is especially true for computing the kinematic integrals using Methods 1 and 3. In these methods we obtained these integrals by computing the integrals in Eq. (1) numerically by using the binned velocity maps derived from the CALIFA datacubes. We estimated the uncertainties of these integrals by computing the dispersion of the radial kinematic integrals at large radius (beyond the bar radius). For most of our galaxies (10/15), these new uncertainties will increase by less than 25$\%$ the reported uncertainties in Table \ref{tb4}. Exceptions are the galaxies NGC5406, NGC6497, NGC6945, UGC03253, and UGC12185. For these galaxies the variation on the errors is between 30\% and 35$\%$. 

\subsection{Ultrafast bars}

Theoretically, the bar cannot end beyond the corotation radius of the galaxies \citep[see][]{contopoulos1980,athanassoula1992}, because the main family of orbits constituting the bar (the $x_{1}$ orbits) is unstable beyond CR. This implies that ${\cal R}$ cannot take values that are less than unity. One of our galaxies (NGC 6497) 
 shows ${\cal R} < 1$  within the errors (1$\sigma$) for our four different measurements.  In addition, NGC 0036 and NGC 5205 show ${\cal R}$ values less than unity within the errors for three of the methods. A visual inspection of these galaxies in Fig. \ref{f1} reveals no special features. There are several ways  to explain these ${\cal R}<1$ values: i) uncertainties associated to the bar and/or corotation radii determination; ii) multiple pattern speeds present in the galaxy; iii) some other reasons for the TW not being applicable to these galaxies. However, we cannot rule out the possibility that they might really be ultrafast bars. 

The parameter ${\cal R}$ is a ratio between the corotation radius and the bar radius. Thus, ${\cal R}$ can be less than one because the bar radius is overestimated and/or the corotation radius is underestimated. We discarded the case of large bars after a careful visual inspection of the images. Moreover, these three galaxies show values of ${\cal R}$ below one even if we use the lowest values of their bar radius given in Table \ref{tb2}. We computed  $R_{\rm CR}$ using the asymptotic flat rotation velocity obtained from the asymmetric drift correction to the LOS velocities of the galaxies. This was obtained after some assumptions on the velocity curve of the galaxies (see Sect. 5.2). We have found the maximum circular velocity measured from HI data for one of the ultrafast bars \citep[NGC0036][]{theureau1998}. For this galaxy,  we have $V_{c,HI}/V_{c,flat}=1.14$. This means that if assuming $V_{c,HI}$ the corotation radius and $\cal R$ would be $14\%$ greater than the values given in Tables \ref{tb3} and \ref{tb4}. Thus, if using $V_{c,HI}$ we would have $\cal R$$ _1=1.0$, $\cal R$$_{2}=0.7$, $\cal R$$_{3}=0.7$, and $\cal R$$_{4}=0.7$. Assuming the same errors as given in Table \ref{tb4}, this galaxy would only be ultrafast for two measurements. This means that the reason that this galaxy is ultrafast could be related to uncertainties associated to the corotation radius determination.

 We have not found in the literature any HI velocity measurements for the two other ultrafast bars (NGC5205 and NGC6497). However, we did recompute the $R_{\rm CR}$  for NGC5205 and NGC6497 using the angular velocity curve obtained from the gas velocity curve along the major axis of these two galaxies \citep[see for details][]{garcialorenzo2015}. In these two cases, the new values of ${\cal R}$ were the same as those obtained from the stellar kinematics. Thus we cannot discard that these two galaxies ${\cal R}$ might be smaller than one within the errors.

The observed distribution of ${\cal R}$ values is given by ${\cal R}_{obs}={\cal R}_{int} \bigotimes {\cal R}_{err}$, where ${\cal R}_{obs}$, ${\cal R}_{int}$, and ${\cal R}_{err}$ are the observed, intrinsic, and error distributions, respectively. To avoid ultrafast bars, we can assume that ${\cal R}_{int}$ is always larger than 1.0. We can ask whether the observed ultrafast bars are only caused by the convolution of the intrinsic $\cal{R}$ distribution, where the observational errors or additional errors in the TW method are indeed needed.

The distribution of ${\cal R}_{1}$ values in our sample of galaxies peaks at ${\cal R}=1.3$ with 37$\%$ of the galaxies with ${\cal R}_{1}<1$. We assume that ${\cal R}_{int}$ is centred at 1.3 with dispersion $\sigma_{int}=0.1$, in order to avoid galaxies with ${\cal R}_{int}<1$. The convolution of ${\cal R}_{int}$ with the observational errors produces a distribution where $\sigma=0.3$ and 15$\%$ of the galaxies show ${\cal R}<1$. Thus, the full fraction of observed ultrafast bars cannot be explained by the convolution of ${\cal R}_{int}$ with the observational errors. We have to convolve ${\cal R}_{int}$ with a distribution ${\cal R}_{err}$ with $\sigma=0.5$ in order to obtain  the 37$\%$ of ultrafast bars. The dispersion of the new error distribution was obtained by adding the dispersion of the observational errors in quadrature plus the dispersion of another distribution of additional errors in the TW method.  The additional errors in the TW turned out to have a dispersion that is 1.3 greater than the dispersion of the distribution of the observed ones. The large extra errors that are needed make it unlikely that ultrafast bars are just a consequence of new errors not being taken into account  in the TW method. Similar results were obtained for ${\cal R}_{2}, {\cal R}_{3},$ and ${\cal R}_{4}$ distributions.

The presence of dust lanes within the bar region can affect the value of the pattern speed of bars determined by the TW method \citep[see][]{gerssen2007}. This variation depends on the angle that the slits of the TW method cut to the bar ($\Delta PA_{bar}$). Thus, dust lanes on the leading edges of a bar tend to increase the TW-derived value of $\Omega_{b}$ when $\Delta PA >0$, and decrease it when $\Delta PA_{bar} <0$. In addition, this variation depends on the amount of extinction. Thus, the variation in the TW-derived pattern speed of the bar is about 8~\% to 25~\%  for galaxies with $A_{V} \approx 3$ in the bar dust lines. This variation increases to 20~\%\ 40~\%\  for $A_{V} \approx 8$. Two of the ultrafast bars, NGC0036 and NGC5205, show $\Delta PA<0$. This means that for these two galaxies, the presence of dust lanes within the bar radius should lead to lower TW values of $\Omega_{b}$ than would the real one. Therefore the TW-derived value of $\cal{R}$ would be greater than the real ones owing to the presence of dust lanes. We can conclude that the low values of $\cal{R}$ obtained for these galaxies are not due to the strong dust lanes within the bar regions. In contrast,  $\Delta PA_{bar}>0$ for NGC6497. Thus, according to Gerssen \& Debattista (2007), the TW method will show  higher values of $\Omega_{b}$ and lower values of $\cal{R}$ than the real ones owing to dust lanes within the bar region. For this galaxy we obtained the extinction map from the fit of its stellar population \citep[see][]{perez2013,gonzalezdelgado2014,cidfernandes2013}. The mean and maximum $A_{V}$ values within the bar region of NGC6947 are 0.16 and 0.66, respectively. Thus, we expect maximum uncertainties of 25$\%$ in the value of $\cal{R}$ (see Debattista \& Gerssen 2007). These uncertainties are not enough to give ${\cal R} > 1$ within the errors (see Table 4).  

Ultra-fast bars have also  been observed in other studies of barred galaxies. In particular, some of the barred galaxies in the sample by Buta \& Zhang (2009)  show ${\cal R}<1$. These authors argue that some of them could be true ultrafast bars and not artefacts due to wrong measurements. Other TW measurements of bar pattern speed have also shown ${\cal R}<1$  \cite[see][]{corsini2007}. The results for the mean value of ${\cal R}$ and the dependence of ${\cal R}$ on Hubble type presented here do not change if we exclude these two galaxies. 

We cannot discard that the TW method cannot be applied for the two ultrafast bars owing to these galaxies not following some of the assumptions of the method. Further research on this direction is therefore needed.

\subsection{Astrophysical implications of fast bars}

Numerical simulations have shown that the bar-pattern speed depends on the bar formation and evolution. If a bar forms by a global bar instability, then it tends to be a fast bar \citep[e.g.,][]{sellwood1981}. Nevertheless, some simulations have shown that bars formed by a gradual bar growth \citep[][]{lyndenbell1979}, by interactions with other galaxies \citep[][]{miwa1998}, or when the initial bulge-to-disc ratio is low \citep[][]{combes1993} can form as slow bars. In particular, Combes \& Elmegreen (1993)  show that late-type galaxies have a low-mass concentration at $\Omega - \kappa/2$ (with $\kappa$ the epiciclic frequency) so that at the beginning of bar formation the CR is far out in the disc, beyond the disc scale length. In contrast, the bar length is determined by the disc scale length. Then, the bar starts out as slow. This does not happen for early-type galaxies where the CR is shorter and determines the bar length, so that the bar is fast from the beginning.  According to these simulations, and assuming no change in the pattern speed after bar formation, we should expect that the bar pattern speeds of early- and late-type barred galaxies should be different, because they are slower for late-type ones. 

Previous observational works show contradictory results. Aguerri et al. (1998) used a sample of ten barred galaxies to find a hint of variation in the bar-pattern speed with the morphological type. In the same direction, Rautiainen et al. (2008)  conclude that late-type galaxies have slow bars, while early-type galaxies do not. In contrast, Buta \& Zhang (2009) find that there is no dependence of ${\cal R}$ on the morphological type. Before the present work, only the pattern speed of three late-type galaxies was obtained by the TW method \citep[e.g.,][]{gerssen2003,treuthardt2007}. These three galaxies turned out to be compatible with fast rotators. Our result confirms the previous findings, and we clearly show no trend of ${\cal R}$ with the Hubble type. Bars located in early- and late-type galaxies are compatible with being fast bars. This result conflicts with the different origins of early and late-type bars proposed in the Combes \& Elmegreen (1993) simulations.

The pattern speed of the bars not only depends on their initial values, but the subsequent bar evolution also can largely change the initial pattern speed. In this sense, several numerical simulations have shown that the interaction between the bar and the dark matter halo of the galaxy produce an exchange of angular momentum at the galaxy resonances that can change the bar-pattern speed \citep[see][]{debattista1998,debattista2000,athanassoula2003,martinezvalpuesta2006}. Our measurements of fast bars across the Hubble sequence indicate that if bars are initially fast, they interchange few angular momentum with the dark matter halo independently of the Hubble type. Therefore bars are fast during their lifetimes \citep[see also][]{perez2012}.

Bar formation and evolution are complex and multiparametric problems involving several galaxy parameters, such as gas, mass fraction, triaxiality of dark matter haloes, and disc velocity dispersion \citep[see][]{athanassoula2003, athanassoula2013}. Our observational results are important for constraining the different formation and evolution scenarios proposed by future numerical simulations.

\section{Conclusions}

We have used integral field spectroscopy data for 15 CALIFA barred  galaxies to determine their bar-pattern speed using the model-independent TW method. Both the FOV and the spatial resolution of the CALIFA data make this dataset appropirate for studies of extended disc galaxies, as proposed here. Integral-field data have the advange of solving problems such as the centring and the disc PA uncertainties that can affect the measured value of the bar-pattern speed by the TW method when long-slit spectroscopic observations are used.

The sample of galaxies selected for this study spans a wide range of morphological types from SB0 to SBbc with most of them being spiral galaxies (SBb-SBbc). This sample fills the gap present in the literature around spiral galaxies for which the TW has not been extensively applied. 

For each galaxy we determined their bar pattern speed using four different approaches to the TW method. The kinematic integrals of the TW were obtained directly from the stellar velocity maps, after summing up the spectra along the corresponding pseudo slits and then measuring the velocity. In addition, we used light and mass as weights of the kinematic and photometric TW integrals. We also obtained the distance-independent quantity ${\cal R}=R_{\rm CR}/a_{\rm b}$ for each galaxy. The corotation radius of the galaxies was approximated by $R_{\rm CR}=V_{\rm c,flat}/\Omega_{\rm b}$, where $V_{\rm c,flat}$ is the asymptotic circular velocity of the galaxies derived after the asymmetric drift correction was applied to the LOS velocity along the galaxy's major axis.

Our CALIFA sample and those galaxies available in the literature with bar pattern-speed measurements by the TW method make a full sample of 32 galaxies from SB0 to SBbc morphological types. This full sample is appropriate to studying the possible dependence of the bar pattern speed on the morphological type.

We found that the mean value of ${\cal R}$ for all galaxies of the sample is in the interval 1.0-1.1. We computed, through Monte Carlo simulations, that we cannot rule out (at 95$\%$ level) the presence of a fast bar ($R<1.4$) in any of our galaxies. The mean value of ${\cal R}$ found for our sample agrees with hydrodynamical numerical simulations of bar galaxies that predict ${\cal R}$ in the interval $[1.0,1.4]$. We did not find any trend of ${\cal R}$ with the Hubble type, so early and late types are both compatible with being fast bars. 

In the near future, we will increase the number of pattern-speed measurements by the TW method using more barred galaxies observed in the CALIFA sample. This will be useful for enlarging the sample of late-type galaxies, decreasing the uncertainties in the mean values of ${\cal R}$, and analysing eventual trends in the bar-pattern speed with other galaxy parameters.

\begin{acknowledgements}
JALA have been partly funded by the Spanish Ministry for Science, project AYA2013-43188-P. R.A. Marino is funded by the Spanish programme of the International Campus of Excellence Moncloa (CEI). I.M. acknowledges the financial support from the Spanish grant AYA2010-15169 and from the Junta de Andalucia through TIC-114 and the Excellence Project P08-TIC-03531. RGD and EP have been partly funded by Spanish grant AYA2010-1581. J.I.P. acknowledges financial support from the MINECO AYA2010-21887-C04-01 grant and from Junta de Andalucía Excellence Project PEX2011-FQM7058. This study makes uses of the data provided by the Calar Alto Legacy Integral Field Area (CALIFA) survey (http://www.califa.caha.es). Based on observations collected at the Centro Astron\'omico Hispano Alem\'an (CAHA) at Calar Alto, operated jointly by the Max-Planck-Institut fur Astronomie and the Instituto de Astrof\'isica de Andaluc\'ia (CSIC). CALIFA is the first legacy survey being performed at Calar Alto.The CALIFA collaboration would like to thank the IAA-CSIC and MPIA-MPG as major partners of the observatory, and CAHA itself for the unique access to telescope time and support in manpower and infrastructures. The CALIFA collaboration also thanks the CAHA staff for their dedication to this project. RGD, EP, RGB, and CCF want to acknowledge the financial support from AYA2010-15081.
\end{acknowledgements}

\end{document}